# Quantum Mechanics from Symmetry

**Roger A. Hegstrom[1] and Alexandra J. MacDermott [2]**

[1] Wake Forest University, Winston-Salem, NC, USA; hegstrom@wfu.edu
[2] University of Houston-Clear Lake, Houston, TX, USA; macdermott@uhcl.edu

## Abstract
Several recent studies have suggested that incompatible variables, which play an essential role in quantum mechanics (QM), are, somewhat surprisingly, not necessarily unique to QM. To investigate this possibility and obtain a better understanding of two central postulates of QM, namely the "commutator postulate" $[\hat{x}, \hat{p}_x] = i\hbar$ and the "Born postulate" $\mathrm{P}_\psi(o_i) = |\langle o_i|\psi\rangle|^2$, we introduce a classical probabilistic theoretical framework which is more general than QM and contains QM as a special case. We call this framework the General Incompatible Variables (GIV) theory, and we show that not only QM systems but also *any* probabilistic systems (classical or quantal) that possess incompatible variables will exhibit the "quantal" properties of uncertainty and interference, and we illustrate this with a roulette-like classical system (which we call the "Arrow") that shows precisely these properties. We show that QM emerges naturally from the GIV framework when the fundamental variables are taken to be symmetries, so that the incompatibility of the QM variables is actually just the incompatibility of the corresponding symmetries (or their generators). Specifically, when the variables are taken to be the elements of the Poincaré group (the symmetry of spacetime) we find that the "commutator postulate" and the "Born postulate" follow automatically, and are therefore no longer postulates. QM thus emerges as a special case of GIV theories for which the variables are symmetries.

## 1 Introduction

In this 2025 "year of the quantum", quantum mechanics (QM) is a century old, and yet, despite its agreement with experiment to high precision, its physical interpretation remains controversial, leading some to suggest that QM is incomplete. In this paper we take the view that an important first step towards a better understanding of QM is the recognition that incompatible variables are *not unique to QM*, and that *any* system with incompatible variables will show *uncertainty* and *interference*. Incompatible variables, along with the concomitant uncertainty and interference, have long been considered the exclusive hallmark of QM, but examples of purely classical, non-quantal systems that exhibit this supposedly "quantal" behaviour have now been found, such as Kirkpatrick's card game models [1] and the Hegstrom-Adshead model of MBCT [2].

The late Steven Weinberg suggested [3] it would be useful, in order to test QM, to find a larger, more general theory in which QM appears as a special case. We have done precisely that: because QM includes incompatible variables, we have constructed a general framework to describe *any* system with incompatible variables, in order to identify the specific features that define a genuinely quantal system. We call our theory the *general incompatible variables* (GIV) theory, and in this paper we show that what distinguishes QM from other GIV theories is *symmetry*: at the most fundamental level, the incompatible variables of QM are *incompatible symmetries* or the generators of incompatible symmetries.

Different textbooks provide varying numbers of postulates of quantum mechanics, which typically boil down to a technical preamble, to the effect that observables are represented by Hilbert space operators $\hat{O}$ that are not only linear but also Hermitian (to give orthogonal eigenstates and real eigenvalues), and that the eigenvalues $o_i$ of these operators represent the possible outcomes of measurements,

$$\hat{O}|o_i\rangle = o_i|o_i\rangle \qquad (1)$$

followed by two central postulates, often described as "underivable". The first postulate, which we will refer to as the "commutator postulate", is that the operators $\hat{O}$ are chosen to obey commutation relations such as the familiar

$$[\hat{x}, \hat{p}_x] = i\hbar \qquad \text{"commutator postulate"} \qquad (2)$$

for position and momentum. The second postulate, commonly known as the "Born postulate", or "Born Rule", is that the probability of each outcome $o_i$ is given by the square of the overlap of the state vector $|\psi\rangle$ with the eigenstate $|o_i\rangle$ corresponding to that outcome,

$$\mathrm{P}_{\psi}(o_i) = |\langle o_i|\psi\rangle|^2 = \cos^2 \omega_{\psi o_i} \qquad \text{"Born postulate"} \qquad (3)$$

where $\mathrm{P}_{\psi}(o_i)$ is the probability of obtaining the outcome $o_i$ when the system is in state $|\psi\rangle$, and $\omega_{\psi o_i}$ is an angle relating the normalized state vectors $|\psi\rangle$ and $|o_i\rangle$ in a single Hilbert space containing all possible state vectors for the system of interest.

It is the "commutator postulate", reflecting the incompatibility of the fundamental variables of QM, that gives rise to the supposedly "quantal" phenomena of uncertainty and interference. We emphasize that we are using here the terms "compatible" and "incompatible" in their familiar, dictionary-definition sense, which in QM takes a somewhat complicated form, namely that two variables A and B are compatible if the corresponding operators $\hat{A}$ and $\hat{B}$ commute, which then, according to a well-known theorem of QM [4], implies the alternative and equivalent definition that A and B are compatible if and only if there exists a complete orthonormal set of common eigenstates $(a_i, b_j)$ of A and B (see also Appendix A of reference [2]). Conversely, the variables A and B are incompatible if the corresponding operators do *not* commute and thus do *not* have a complete set of common eigenstates. We emphasize this because some authors – such as Hughes [5] and Kirkpatrick [1], as we see later – use a more restricted meaning of incompatibility that not only requires the variables to be not compatible (with non-commuting operators and no complete set of common eigenstates) but also, in addition, requires them to be representable in one and the same Hilbert space.

Frank Wilczek [6] has lamented the fact that QM does not appear to have a guiding principle based on symmetry, unlike relativity (based on the equivalence of different inertial frames) or gauge theory (based on the equivalence of different potentials). But in fact an attempt *has* been made to derive QM from symmetry in a 1995 paper [7], by Aage Bohr (the late son of Niels Bohr) and Ole Ulfbeck, that deserves to be much more widely known. Bohr & Ulfbeck aimed to show that, contrary to what we all tell our undergraduates, the two central postulates identified above – the "commutator postulate" and the "Born postulate" – have little to do with either wave character or quantization, but instead result from the Poincaré symmetry of spacetime. Their account of the derivation of the "commutator postulate" from symmetry represents a major insight, as we discuss in Section 3. But their attempt to derive the "Born postulate" $\mathrm{P}_{\psi}(o_i) = |\langle o_i|\psi\rangle|^2$ from symmetry does not work because it tacitly assumes that $|\psi\rangle$ and $|o_i\rangle$ are represented in a *single* Hilbert space, but ignores Gleason's theorem [8], which shows that, if $|\psi\rangle$ and $|o_i\rangle$ are represented in a single Hilbert space, the *only* probability expression that works is the Born Rule, symmetry or no symmetry. In fact Weinberg pointed out in 2015 [9] that it is unclear where the "Born postulate" comes from, and that attempts to "derive" it have hitherto been circular. Most of these attempts are based upon deep QM principles, usually in the context of specific interpretations of QM, most notably by Deutsch [10] and Wallace [11], based on the many-worlds interpretation, and by Zurek [12], based on decoherence by entanglement with the environment. But it is less widely known that just as incompatible variables are not unique to QM, the Born Rule is also not unique to QM, since Brumer & Gong [13] have derived an analogue of the Born Rule within purely classical mechanics.

We take the view that the "Born postulate" $\mathrm{P}_{\psi}(o_i) = |\langle o_i|\psi\rangle|^2 = \cos^2\omega_{\psi o_i}$ is not really a "postulate" as such, but rather is best seen as simply a free Pythagorean *construction* (see Appendix 1) for accommodating basic features of classical probability theory in Hilbert spaces. Hilbert spaces are intrinsically Pythagorean in nature [5], so we term the angle $\omega_{\psi o_i}$ relating $|\psi\rangle$ and $|o_i\rangle$ in Hilbert space the *Pythagorean angle*. Although commonly associated with QM, Hilbert spaces are

not unique to QM and can be applied to any probabilistic system, including the familiar coin toss [3]. One can always *choose* to use the Born Rule as a construction (since it relies only on the truth of Pythagoras' theorem), *provided* that $|\psi\rangle$ and $|o_i\rangle$ are vectors in the same Hilbert space and that the eigenvectors $|o_i\rangle$ form a complete orthonormal set in that space. But Gleason's theorem goes further, proving that the Born Rule is not merely a convenient choice of probability rule but actually the *only* possible choice if all vectors describing the system belong to a single Hilbert space (see Appendix 1).

So what distinguishes QM from classical systems with incompatible variables? And what kind of formalism do non-QM GIV theories use? In QM, all possible state vectors for a system are represented in a single Hilbert space, and the probability of obtaining a particular result when measuring any one of the variables is given by a single Born Rule, as in equation (3). But it was noticed by Hughes [5] that if two variables A and B are *not compatible* (i.e. incompatible in the familiar dictionary-definition sense mentioned earlier) they *cannot in general* (except in special cases) be represented together in just one Hilbert space. This is because *both* sets of eigenvectors (those of A and those of B) would have to form complete orthonormal sets in that Hilbert space, and this is *not in general possible* for *incompatible* variables, because the condition of orthogonality for these vectors imposes severe and unnecessary restrictions on the form of the probability functions that would arbitrarily exclude many possible GIV systems, as we demonstrate in Section 4. To get around this problem, Hughes introduced the idea of using a separate Hilbert space for each of the non-compatible variables [5], but did not develop this approach any further because QM has what he described as the "remarkable feature" of containing variables that are not compatible and yet *can* be represented in the same Hilbert space. He therefore chose to define "incompatible" variables as those that are not only *not compatible* but also representable in a single Hilbert space – and Kirkpatrick's work [1] also tacitly assumes a single Hilbert space. We prefer to describe variables that are not compatible and also representable in a single Hilbert space as "quantum incompatible", while using the terms "ordinary incompatible", "general incompatible", or simply "incompatible", for the familiar meaning of just *not compatible* (but not necessarily representable in a single Hilbert space).

In this paper we provide an explanation, in terms of symmetry, for the "remarkable feature" of QM that its variables *can* be represented in a single Hilbert space, despite being incompatible in the ordinary sense. Following Hughes [5], our GIV theory uses many Hilbert spaces, one for each incompatible variable: the eigenvectors of A are orthogonal in Hilbert space $\mathcal{H}_A$ (in which the eigenvectors of A form the axes), but the vectors representing the eigenstates of B are *not in general* orthogonal in Hilbert space $\mathcal{H}_A$; similarly the eigenvectors of B are orthogonal in Hilbert space $\mathcal{H}_B$ (in which the eigenvectors of B form the axes), but the vectors representing the eigenstates of A are *not in general* orthogonal in $\mathcal{H}_B$. Each of our multiple Hilbert spaces uses what we call a *restricted* version of the Born Rule $P_\psi(o_i) = |\langle o_i|\psi\rangle|^2$ for the probability of getting the outcome $o_i$ when making a measurement on the state $|\psi\rangle$. In the restricted Born Rule, $|\psi\rangle$ can be any vector in the Hilbert space, but $|o_i\rangle$ is restricted to being one of the fixed set of orthonormal vectors defining the axes of that particular Hilbert space; this is in contrast to the usual version of the Born Rule, in which not only $|\psi\rangle$ but also $|o_i\rangle$ can be *any* vector within the Hilbert space. So probabilities of A outcomes can in general only be calculated in Hilbert space $\mathcal{H}_A$, and similarly probabilities of B outcomes can only be calculated in Hilbert space $\mathcal{H}_B$, if the variables A and B are incompatible.

So why is QM unique among GIV theories in using only one Hilbert space? Is there something special about the incompatible variables of QM? And what is the source of their incompatibility? The source of the incompatibility of variables in certain classical systems is typically rather mundane and obvious, such as the fact that an arrow cannot be pointing in two directions at the same time, as we discuss in Section 4. But the source of the incompatibility of the QM variables is much more profound: the QM variables are *symmetries*, and *the incompatibility of the QM variables is actually the incompatibility of the corresponding symmetries*, as we discuss in Section 2. This kind of incompatibility is easy to visualize and understand: for example, an object having cylindrical symmetry as its highest symmetry cannot be cylindrically symmetric about both the *x*-axis and the *y*-axis at the same time, leading to the non-commutation $[\hat{J}_x, \hat{J}_y] = i\hbar\hat{J}_z$ of the angular momentum operators $\hat{J}_x$ and $\hat{J}_y$ that are the generators of rotations about the respective axes, and to the non-existence of a complete set of simultaneous eigenstates for $\hat{J}_x$ and $\hat{J}_y$. Similarly, as pointed out by Bohr & Ulfbeck [7], a space-time object cannot have both Lorentz symmetry and translational symmetry in the *x*-direction at the same time, leading [7,14] to the non-commutation $[\hat{K}_x, \hat{p}_x] = \frac{i}{c}\hat{H}$ of the generator $\hat{K}_x$ of Lorentz transformations and the corresponding momentum $\hat{p}_x$ (generator of translations), which in the non-relativistic regime becomes $[\hat{x}, \hat{p}_x] = i\hbar$ where $\hat{x} = (\hbar/mc)\hat{K}_x$, as we discuss in Section 3. We see that $\hat{K}_x$ plays a double role: it not only acts as the generator of Lorentz transformations about a spacetime point but also provides a deeper, more fundamental definition of the position operator $\hat{x}$.

The fact that the eigenstates of incompatible variables cannot *in general* be represented together in just one single Hilbert space has, to our knowledge, been noticed only by Hughes [5]. But what has *not*, to our knowledge, been noticed until now is that the eigenstates of incompatible variables *can* be represented in the same Hilbert space (and indeed *must* be representable in the same Hilbert space) if the incompatible variables are incompatible *symmetries* (as in QM)!

We show in Section 5 that if the incompatible variables are symmetries, the many Hilbert space formalism of GIV theories can then be collapsed into the familiar QM formalism of a single Hilbert space with just a single, unrestricted Born Rule. And why is it uniquely symmetry that allows the many Hilbert spaces to be combined into one? We argue that when the fundamental variables are symmetries, as in QM, the many Hilbert spaces of GIV theory are then related to one another by *rigid* transformations (as opposed to non-rigid transformations in the general case), which ensures that the vectors representing the eigenvectors of a variable B are orthogonal not only in Hilbert space $\mathcal{H}_B$, but also in Hilbert space $\mathcal{H}_A$, enabling the many Hilbert spaces to be superimposed and amalgamated into one.

So symmetry is behind the simple, single Hilbert space formalism of QM. The role of symmetry in the "commutator postulate" is also clear, as shown by Bohr & Ulfbeck and others and summarized in Section 3. But does symmetry also play a role in the "Born postulate", as Bohr & Ulfbeck [7] attempted to show? In this paper we show that the Born Rule is not *primarily* a consequence of symmetry, since its unrestricted form for a single-Hilbert-space theory is required by Gleason's theorem [5, 8] and its restricted form in a many-Hilbert-space theory is a free Pythagorean construction. But when the incompatible variables are *incompatible symmetries*, the many-Hilbert-space description of GIV theory, in which only restricted forms of the Born Rule apply, becomes physically equivalent to a *single* Hilbert space description, in which the unrestricted Born Rule applies. We have thus proved that although the Born Rule is not in itself derived from symmetry, it *is* symmetry that allows the *restricted* Born Rule to take on its familiar *unrestricted* form, in agreement with Gleason's theorem.

In Section 2 we clarify what we mean when we say that the incompatible *variables* of QM are actually incompatible *symmetries*, and in Section 3 we review the role of symmetry in the "commutator postulate". We then turn to the "Born postulate", starting in Section 4 with a simple classical system that we call the "Arrow" as a specific example to illustrate the use of many Hilbert spaces for a system with incompatible variables and to highlight the role of symmetry in enabling a single Hilbert space (and thence the familiar *unrestricted* Born Rule) to be used. In Section 5 we outline how our treatment of the Arrow can be generalized in our GIV theory. In Section 6 we summarize our discoveries about the role of symmetry in the "Born postulate", which now complete the work, started by Bohr & Ulfbeck, of showing that both the "commutator postulate" *and* the "Born postulate" do truly come from symmetry. Finally, in Section 7 we use

GIV theory to demonstrate that *any* system, classical or quantal, wave-like or not, will show *uncertainty* and *interference* in its probability patterns if it has *incompatible variables*. We find that symmetries in our "Arrow" system lead to an enhanced *impression* of wave character in its interference patterns (even though there are no *physical* waves in this purely classical system), in direct analogy with symmetries present in the double-slit system that also lead to an enhanced impression of wave character.

## 2 Variables as symmetries

Before proceeding further, we need to clearly establish what we mean by the incompatible *variables* of QM being incompatible *symmetries*. In QM the symmetries of spacetime (homogeneity, isotropy, Lorentz invariance) play *two* roles. The symmetries' most obvious role – as the operators that transform the states under symmetry transformations – is the one that is usually emphasized. But Bohr & Ulfbeck gave more importance to the symmetries' other role as the fundamental variables of a QM system. They argued that this second, less appreciated role should actually be considered their primary role, because it is the incompatibility of certain *symmetries* that makes certain *variables* incompatible, leading to the "commutator postulate" and the resulting Heisenberg Uncertainty Principle that is at the heart of QM – hence the title "Primary manifestation of symmetry. Origin of quantal indeterminacy" of their landmark 1995 paper [7].

"Variables" have *values*: the "value states" (eigenstates) of a variable A are represented in Hilbert space as the basis vectors $|a_i\rangle$ satisfying the eigenvalue equation

$$\hat{A}|a_i\rangle = a_i|a_i\rangle \tag{4}$$

where $a_i$ is the value of A. And since *symmetry* operations by definition leave an object unchanged, we can certainly appreciate that a *symmetry* operator $\hat{A}$ may also have eigen *states* that it leaves unchanged upon measurement as in equation (4). But how can a *symmetry* operator have eigen *values*? The notion of symmetries having a "value" becomes clearer when we realize that most symmetries in nature are *continuous* symmetries (e.g. rotations through any of a *continuous* range of angles). Continuous symmetries are dealt with by Lie groups, in which the symmetry operators $\hat{A}$ have the general form

$$\hat{A}(\alpha) = e^{i\alpha\hat{G}} \tag{5}$$

where $\alpha$ is a continuous parameter (e.g. the angle through which a rotation takes place) and $\hat{G}$ is the *generator* of the symmetry operator. Note that the operator $\hat{A}$ is unitary ($\hat{A}\hat{A}^\dagger = 1$), provided the generator is Hermitian ($\hat{G} = \hat{G}^\dagger$). Since the effect of an exponential operator such as $e^{i\alpha\hat{G}}$ is defined in terms of the series expansion of the exponential

$$e^{i\alpha\hat{G}} = \sum_{n=0}^{\infty} \frac{(i\alpha\hat{G})^n}{n!} \quad (6)$$

it is clear that the eigenstates of $e^{i\alpha\hat{G}}$ are in fact the eigenstates $|g_i\rangle$ of the generator $\hat{G}$:

$$\hat{G}\,|g_i\rangle = g_i|g_i\rangle \quad (7)$$

$$e^{i\alpha\hat{G}}\,|g_i\rangle = e^{i\alpha g_i}|g_i\rangle \quad (8)$$

Since these eigenstates $|g_i\rangle$ of $\hat{G}$ are also eigenstates of $\hat{A}(\alpha)$, we can choose to label them either with the eigenvalues $a_i(\alpha) = e^{i\alpha g_i}$ of $\hat{A}(\alpha)$ or with the eigenvalues $g_i$ of $\hat{G}$ (or even with both), so equation (8) can be written

$$\hat{A}(\alpha)|a_i(\alpha)\rangle = a_i(\alpha)|a_i(\alpha)\rangle = e^{i\alpha g_i}|a_i(\alpha)\rangle = e^{i\alpha g_i}|g_i\rangle \quad (9)$$

The "value" of the symmetry "variable" A(α) in the state $(a_i(\alpha))$ is thus

$$a_i(\alpha) = e^{i\alpha g_i} \quad (10)$$

It may seem strange to think of symmetries as having "values", but in QM the directly measured variables are nearly always the *generators* of symmetry transformations. Examples include angular momentum $\hat{J}_z$, the generator of rotations $\hat{R}^z(\phi) = e^{-(i/\hbar)\phi\hat{J}_z}$ through an angle $\phi$ about the *z*-axis, and linear momentum $\hat{p}_x$, the generator of translations $\hat{T}^x(a) = e^{-(i/\hbar)a\hat{p}_x}$ through a distance $a$ along the *x*-axis. Since the eigenvalues $g_i$ of the measured variables are so closely related to the eigenvalues $a_i(\alpha) = e^{i\alpha g_i}$ of the symmetry operators, we can equally well consider measurements of the directly measured variables to be measurements of the corresponding symmetries: once we measure $g_i$, we also know the value $a_i(\alpha) = e^{i\alpha g_i}$ of the symmetry for any given $\alpha$.

It is important to recognize that the eigenstates $|a_i(\alpha)\rangle$ of $\hat{A}(\alpha)$ and the eigenstates $|g_i\rangle$ of $\hat{G}$ are one and the same, and do not depend on $\alpha$. The parameter $\alpha$ is a parameter only of the *operator* $\hat{A}(\alpha)$ and its eigenvalue $a_i(\alpha)$ (indicating the angle rotated through, or the distance translated through), but is not a parameter of the eigenstates. We therefore construct our Hilbert space using just the basis vectors $|g_i\rangle$ satisfying

$$\hat{A}(\alpha)|g_i\rangle = a_i(\alpha)|g_i\rangle \quad (11)$$

because then there is no need for separate basis vectors for each value of $\alpha$. Since the basis vectors $|g_i\rangle$ are the same for all $\alpha$, we can simplify the notation by dropping the dependence on $\alpha$ and writing equation (11) as

$$\hat{A}|a_i\rangle = a_i|a_i\rangle \quad (12)$$

as in equation (4). This notation has the advantage of being more general: $\hat{A}$ could now represent the generator $\hat{G}$ of $\hat{A}(\alpha)$, or it could represent any one of the infinite number of operators $\hat{A}(\alpha)$ in the Lie groups relevant to QM, or it could be a more general symmetry operator in any type of group in a General Incompatible

Variables (GIV) theory. We will therefore use equations such as (12) when we introduce our GIV theory in Section 5.

We achieve a simplification if we consider the generator to be the fundamental variable and the symmetry as a *derived* variable, because the number of generators is generally much less than the (infinite) number of group elements in a continuous group. Since for a Lie group both the symmetries and their generators are variables, and these are simply related, as in equations (5) and (10), we can use either or both in any interpretive discussions, and we will see that in QM it is often more useful to consider measurements of the directly measured variables (the generators) as measurements of the corresponding symmetries.

We thus see that when the variables are symmetries (or generators thereof), *incompatible variables* are *incompatible symmetries*. And incompatible symmetries are simply symmetries that cannot be possessed simultaneously by a system. For example, consider a system the highest symmetry of which is cylindrical symmetry about the $x$-axis: such a system cannot also be cylindrically symmetric about the $y$-axis. The incompatibility of these symmetries is reflected in the well-known non-commutation of the generators $\hat{J}_x$ and $\hat{J}_y$ of rotations about the $x$ and $y$ axes respectively, which results in the non-existence of simultaneous eigenstates $|J, M_x, M_y\rangle$.

## 3 The "commutator postulate"

We now summarize how the non-commutation of certain *symmetry* operators leads to the non-commutation of the operators for certain *variables*, as in the "commutator postulate" $[\hat{x}, \hat{p}_x] = i\hbar$. Spacetime has full Poincaré symmetry: it is homogeneous, isotropic, and Lorentz invariant. The symmetry operations of the Poincaré group are therefore space-time translations T, spatial rotations R, and Lorentz "boost" transformations Λ, and the generators of the corresponding symmetry operators are, respectively, the momentum vector $\boldsymbol{p}$ for translations, the angular momentum vector $\boldsymbol{J}$ for spatial rotations, and the "boost" vector $\boldsymbol{K}$ for Lorentz transformations. Some of these symmetry operators do not commute, e.g. rotations about different axes do not commute (as is well-known instinctively even to child solvers of Rubik's cube!), and this results in non-commutation of the *generators* of these rotations, as in the familiar angular momentum commutator $[\hat{J}_x, \hat{J}_y] = i\hbar\hat{J}_z$. It is fairly easy to see that spatial *rotations* do not commute with *translations*, and it turns out that Lorentz transformations also do not commute with translations. This is because Lorentz transformations are also *rotations*, but in spacetime: in a (1 + 1) spacetime the Lorentz transformation for a velocity boost $v$ in the $x$ direction corresponds to a rotation in the $xt$ plane through an imaginary angle $\tanh^{-1}(v/c)$ about the origin $(0,0)$ for a velocity boost $v$ in the $x$ direction (see, for example [15]). The corresponding generators $\hat{K}_x$ for boosts and $\hat{p}_x$ for translations therefore also do not commute:

$$\left[\hat{K}_x, \hat{p}_x\right] = i\frac{\hat{H}}{c} \tag{13}$$

Bohr & Ulfbeck [7] pointed out that a spacetime rotation singles out a spacetime point $(x, t)$ that remains fixed under rotation about that point, and under spatial rotations or translations that point is changed into a different point $(x', t')$, further underlining the non-commutation of Lorentz transformations with both spatial rotations and translations. The generators of Lorentz transformations therefore need to be labelled with the invariant point $(x, t)$ about which the spacetime rotation takes place, and Bohr and Ulfbeck used rotation about the invariant point $(0, t)$ to define the position operator as a function of time as

$$\hat{x}(t) = \frac{\hbar}{mc}\hat{K}_x(0, t) \tag{14}$$

whereupon equation (13) becomes

$$[\hat{x}, \hat{p}_x] = i\hbar\frac{\hat{H}}{mc^2} \tag{15}$$

and in the non-relativistic limit (p << mc) the Hamiltonian becomes equal to the rest mass energy,

$$\hat{H} = mc^2 + \left(\frac{p^2}{2m}\right) \approx mc^2 \tag{16}$$

so that equation (15) becomes

$$[\hat{x}, \hat{p}_x] = i\hbar \tag{17}$$

which is the familiar "commutator postulate" for position and momentum.

Bohr & Ulfbeck make a point of noting in their 1995 paper [7] that the "commutator postulate" (and indeed the entire notion of "position"!) requires the link between space and time introduced by relativity, even for ordinary *non-relativistic* QM. And in fact Melvin had earlier, in his 1960 paper [16], used relativistic quantum field theory (QFT) to derive the expression

$$\hat{x}(t) = \int d^3x \; x\frac{\hat{H}(x,t)}{mc^2} \tag{18}$$

where $\hat{H}(x, t)$ is the Hamiltonian density, constructed from the quantum field operators for the system in the usual way. This expression equates the position operator to what Melvin calls the "centroid of energy". The Bohr & Ulfeck and Melvin expressions (14) and (18) for the position operator $\hat{x}$, which use respectively invariance under symmetry and "centroid of energy" to single out a *point* in space, are equivalent, and both lead to the "commutator postulate" (17). We will discuss this further in a later paper and provide a diagrammatic understanding of the commutator $[\hat{x}, \hat{p}_x]$, analogous to the more familiar understanding of the commutator $\left[\hat{J}_x, \hat{J}_y\right]$. It is worth noting that "centroid of energy" is a remarkably *particle*-like feature to emerge from theories that focus on *waves* (QM) or *fields* (QFT). In Section 7 we show that wave-like interference and uncertainty are not unique to QM, and are not necessarily indicative of actual physical waves, but rather are a general feature of *all* probability theories that have incompatible variables.

In conclusion, once it is recognized that, at the most fundamental level, the QM variables are actually symmetries of the Poincaré group, the "commutator postulate" then automatically follows, because the Poincaré symmetries (and thus the QM variables) must obey the commutation relations of the Poincaré group. So the "commutator postulate" is *not a postulate*, rather it is a straightforward consequence of the symmetry of spacetime! What about the "Born postulate"? Does it too come from symmetry? In the next Section we introduce a simple *classical* system with incompatible variables as a specific example to illustrate how the "Born postulate" does indeed also come from symmetry.

## 4 The "Arrow": a non-QM system with incompatible variables

The coin toss is a non-QM system with a single variable ("sidedness") that has just *two* possible values, heads or tails. Our "Arrow" is a simple non-QM system that is like the coin toss, but with not just one but an infinite number of two-valued variables, all *incompatible* with each other. The Arrow is an ordinary directed object, such as a meter stick, with an arrow-head painted on the "head" end to clearly distinguish it from the "tail" end. The center of the Arrow is fixed at the origin, and its head can be rotated about an axis through the origin, perpendicular to the Arrow, but the Arrow always remains perpendicular to the rotation axis, similar to the rotation of a roulette wheel. We prepare the Arrow in an initial state by orienting it along some initial direction represented by a vector $\vec{\boldsymbol{n}}$ in physical space. To make a measurement of the variable A, we set up a measurement apparatus analogous to a roulette wheel that has only two slots for the ball: the two "slots" for the Arrow are positioned at opposite ends of a vector $\vec{\boldsymbol{a}}$ that passes through the origin at an angle $\theta_{n_+ a_+}$ to $\vec{\boldsymbol{n}}$. The Arrow is then caused (by random "kicks" or "spins") to rotate with damping until it finally settles into one of two possible orientations: (1) Arrow head in $\vec{\boldsymbol{a}}$ direction (A = $a_+$), which we designate state $(a_+)$, or (2) Arrow head in $-\vec{\boldsymbol{a}}$ direction (A = $a_-$), which we designate state $(a_-)$. How this is accomplished is unimportant, but it could be done by attaching a small magnet to the tip of the Arrow and placing two additional magnets at appropriate points along the $\vec{\boldsymbol{a}}$ and $-\vec{\boldsymbol{a}}$ directions. On release from its initial orientation $\vec{\boldsymbol{n}}$, the Arrow rotates and eventually clicks into place with the Arrow head pointing in either the $\vec{\boldsymbol{a}}$ or $-\vec{\boldsymbol{a}}$ "slots", giving one of only two possible outcomes for the variable A, namely $a_+$ or $a_-$; and the probabilities of getting $a_+$ would depend in some way on the initial orientation of $\vec{\boldsymbol{n}}$, i.e. on the angle $\theta_{n_+ a_+}$.

The variable A has only two values, like heads or tails in the coin toss, or spin-up and spin-down in a QM spin ½ system, so it is analogous to a roulette wheel where the ball can only land at either the “12 o’clock” position or the “6 o’clock” position, the two positions being 180° apart, as in Figure 1.

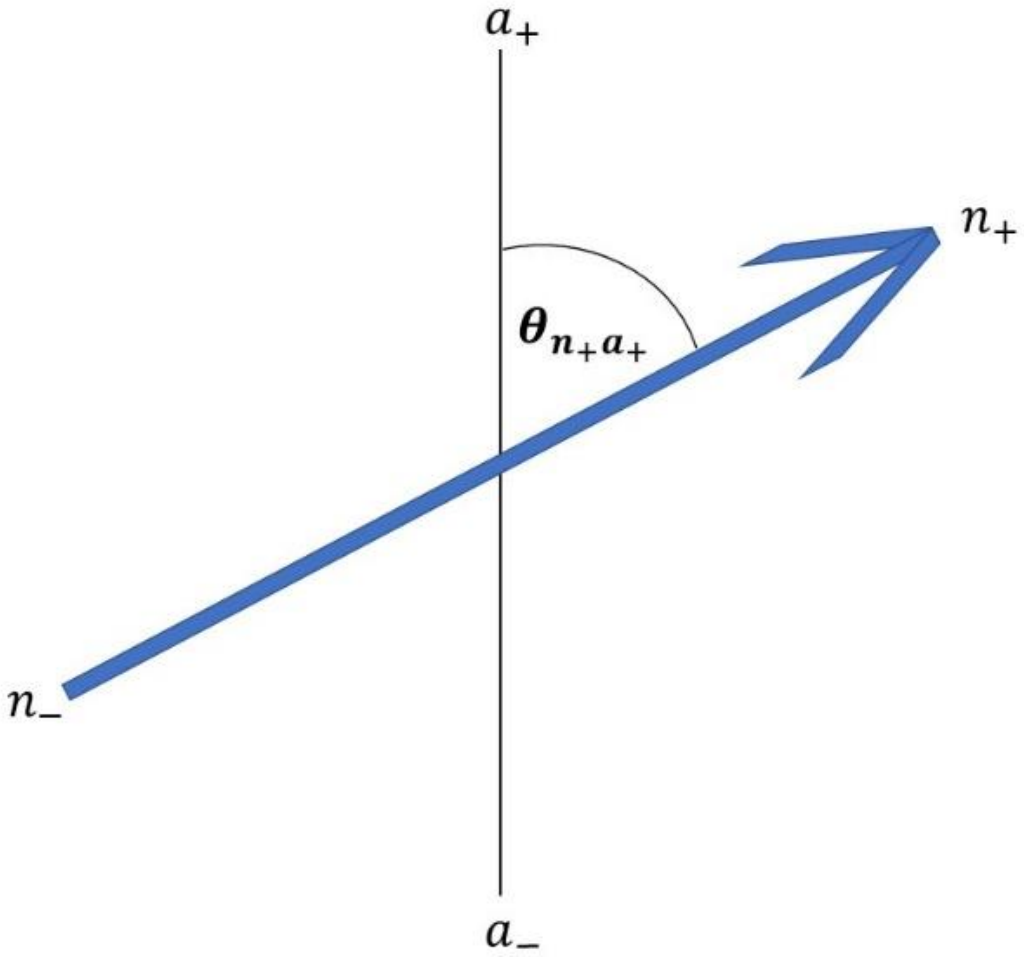

**Figure 1.** Arrow system set up to measure variable A

Having prepared the Arrow in the direction $\vec{\boldsymbol{n}}$, one could alternatively measure another variable B, with values $b_+$ and $b_-$, by instead placing magnets in the “10 o’clock” and “4 o’clock” positions (also 180° apart, Figure 2); and in fact there are an infinite number of variables A, B, C....., one for each direction in the plane of the Arrow.

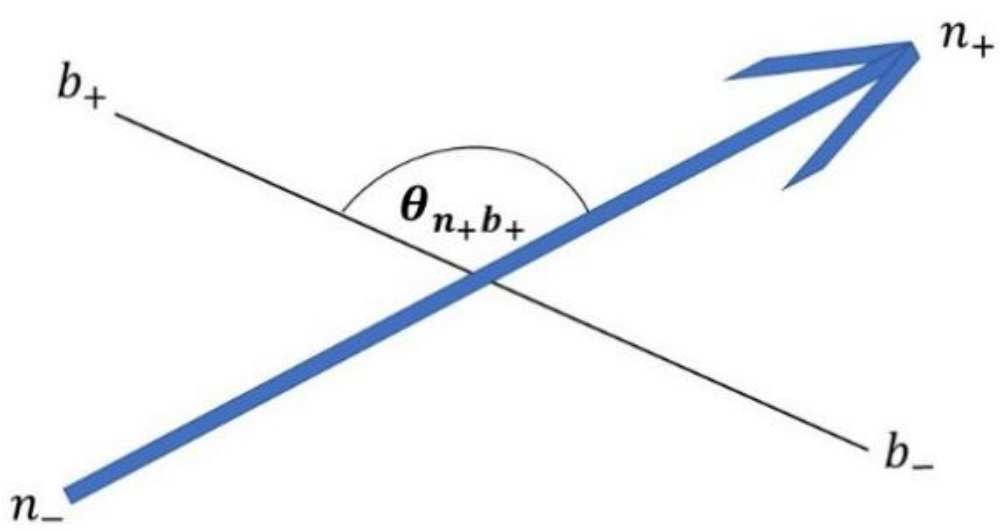

**Figure 2.** Arrow system set up to measure variable B

These variables are all mutually incompatible, as the Arrow ends up pointing in a different direction for each variable, e.g. A and B are incompatible because when the Arrow comes to rest in the $\pm\vec{\boldsymbol{a}}$ direction in the measurement of A, it cannot at the same time be pointing in the $\pm\vec{\boldsymbol{b}}$ direction, as in a measurement of B. There is therefore no state such as $(a_+, b_+)$ for which we could predict the values of both A $(= a_+)$ and B $(= b_+)$ with 100% certainty, in accord with the definition of incompatibility. Measurement of the incompatible variables A and B would in general also involve two different pieces of apparatus, with magnets along different axes, but could be done in a single apparatus using electromagnets in the different directions that could be switched on and off according to which variable was being measured.

If the Arrow is prepared in the state $(n_+)$ (i.e. pointing in the positive $\vec{\boldsymbol{n}}$ direction), the probability of obtaining the outcome $a_+$ upon measuring A will be some function $f_{a_+}$ of the angle $\theta_{n_+a_+}$ between the $n_+$ and $a_+$ directions. We will use the notation $\mathrm{P}_{(n_+)}(a_+)$ for the probability of obtaining the outcome $a_+$ when starting from the state $(n_+)$, so we have

$$\mathrm{P}_{(n_+)}(a_+) = f_{a_+}\left(\theta_{n_+a_+}\right) \tag{19}$$

At this stage, we make *no* assumptions about any symmetries, so there is no reason to assume the same probability function $f$ for measuring the different eigenvalues of A, so we write the probability of obtaining the outcome $a_-$ when starting from the state $(n_+)$ as a *different* function $f_{a_-}$ of the angle $\theta_{n_+a_-}$ between the $n_+$ and $a_-$ directions,

$$\mathrm{P}_{(n_+)}(a_-) = f_{a_-}\left(\theta_{n_+a_-}\right) \tag{20}$$

There is also no reason to assume the same probability functions $f$ for measuring the different variables A and B, so we write the corresponding probabilities for obtaining the respective outcomes $b_+$ and $b_-$ upon measuring B as

$$\mathrm{P}_{(n_+)}(b_+) = f_{b_+}\left(\theta_{n_+b_+}\right) \tag{21a}$$

$$\mathrm{P}_{(n_+)}(b_-) = f_{b_-}\left(\theta_{n_+b_-}\right) \tag{21b}$$

We thus assume for now that $f_{a_+}$, $f_{a_-}$, $f_{b_+}$ and $f_{b_-}$ are all different functions of the respective angles, and symmetry will be introduced later.

So what is the form of the various probability functions $f(\theta)$ ? The different functions $f$ could in general be almost any functions of the respective angles $\theta$, restricted only by the basic requirements of classical probability, namely normalization, exclusivity, value range 0 to 1, and the additional constraints $f(0) = 1$ and $f(\pi) = 0$ arising from the linear geometry of the Arrow and the requirement that a measurement of A performed upon the eigenstate $(a_\pm)$ must give the outcome $a_\pm$ with certainty. Many possible functions satisfy these minimal requirements, including the functions $f(\theta) = 1 - \frac{\theta}{\pi}$, $f(\theta) = 1 - \left(\frac{\theta}{\pi}\right)^2$, $f(\theta) = \cos^2 \tfrac{1}{2}\theta$ and more.

We now consider how to represent the Arrow states in Hilbert space. If the variables A and B were compatible, an appropriate Hilbert space would use the four-dimensional orthonormal basis $\{|a_+, b_+\rangle, |a_+, b_-\rangle, |a_-, b_+\rangle, |a_-, b_-\rangle\}$. But with A and B incompatible, a set of simultaneous eigenstates of A and B such as

this does not exist. So we try instead to use a two-dimensional Hilbert space $\mathcal{H}_A$ with $\{|a_+^A\rangle, |a_-^A\rangle\}$ as an orthonormal basis. But how do we represent the eigenstates of B in Hilbert space $\mathcal{H}_A$? Suppose we prepare the Arrow in the state $(b_+)$ (i.e. pointing in the positive $\vec{\boldsymbol{b}}$ direction in physical space, at an angle $\theta_{b_+a_+}$ to the positive $\vec{\boldsymbol{a}}$ direction and an angle $\theta_{b_+a_-}$ to the negative $\vec{\boldsymbol{a}}$ direction). We can represent the state $(b_+)$ in the Hilbert space $\mathcal{H}_A$ using the Pythagorean construction used by Hughes [5] to obtain

$$|b_+^A\rangle = f_{a_+}^{1/2}(\theta_{b_+a_+})|a_+^A\rangle + f_{a_-}^{1/2}(\theta_{b_+a_-})|a_-^A\rangle \tag{22a}$$

where the probabilities of getting the results $a_+$ and $a_-$ when starting from state $(b_+)$ are $\mathrm{P}_{(b_+)}(a_+) = f_{a_+}(\theta_{b_+a_+})$ and $\mathrm{P}_{(b_+)}(a_-) = f_{a_-}(\theta_{b_+a_-})$ respectively. If instead we prepare the Arrow in the state $(b_-)$ (i.e. pointing in the negative $\vec{\boldsymbol{b}}$ direction, at an angle $\theta_{b_-a_+}$ to the positive $\vec{\boldsymbol{a}}$ direction and an angle $\theta_{b_-a_-}$ to the negative $\vec{\boldsymbol{a}}$ direction), we can similarly represent the state $(b_-)$ in the Hilbert space $\mathcal{H}_A$ as

$$|b_-^A\rangle = -f_{a_+}^{1/2}(\theta_{b_-a_+})|a_+^A\rangle + f_{a_-}^{1/2}(\theta_{b_-a_-})|a_-^A\rangle \tag{22b}$$

where we have introduced the minus sign as a choice of arbitrary phase that will be useful later (to be completely general, the quantities $f_{a_+}$, $f_{a_-}$ in equations (22) should really be complex numbers, but we choose them to be real for the sake of simplicity). Taking into account the normalization of the probabilities, equations (22) can then be written

$$|b_+^A\rangle = f_{a_+}^{1/2}(\theta_{b_+a_+})|a_+^A\rangle + \left(1 - f_{a_+}(\theta_{b_+a_+})\right)^{1/2}|a_-^A\rangle \tag{23a}$$

$$|b_-^A\rangle = -\left(1 - f_{a_-}(\theta_{b_-a_-})\right)^{1/2}|a_+^A\rangle + f_{a_-}^{1/2}(\theta_{b_-a_-})|a_-^A\rangle \tag{23b}$$

using equation (A6) from Appendix 1. As we have seen, the $(a_i)$, $i = +, -$, states of the Arrow are represented in the Hilbert space $\mathcal{H}_A$ by the orthonormal basis $\{|a_+^A\rangle, |a_-^A\rangle\}$ that forms the axes, and the $(b_i)$ states are represented in $\mathcal{H}_A$ by the vectors $|b_\pm^A\rangle$ in equations (23). The $(b_i)$ states would of course be orthogonal in the Hilbert space $\mathcal{H}_B$, in which the orthonormal basis $\{|b_+^B\rangle, |b_-^B\rangle\}$ forms the axes of $\mathcal{H}_B$. But if we evaluate $\langle b_-^A|b_+^A\rangle$ using equations (23), we see that $|b_+^A\rangle$ and $|b_-^A\rangle$ can only be orthogonal in Hilbert space $\mathcal{H}_A$ if

$$\left(1 - f_{a_+}(\theta_{b_+a_+})\right)^{1/2} f_{a_-}^{1/2}(\theta_{b_-a_-}) - \left(1 - f_{a_-}(\theta_{b_-a_-})\right)^{1/2} f_{a_+}^{1/2}(\theta_{b_+a_+}) = 0 \tag{24}$$

which is not in general satisfied unless some of the angles and probability functions $f$ are equal. Attempting to represent the $(b_i)$ states in Hilbert space $\mathcal{H}_A$ and using the Born rule $\mathrm{P}_\psi(b_i) = \left|\langle b_i^A|\psi^A\rangle\right|^2$ will therefore yield incorrect results for the probability of outcomes of measurements of the variable B, because

the basic *exclusivity* rule of probability is violated if the $|b_{\pm}^{A}\rangle$ states are not orthogonal (e.g. $\mathrm{P}_{(b_+)}(b_-) = |\langle b_-^A | b_+^A \rangle|^2 \neq 0$). But *why* are the $(b_i)$ states not in general orthogonal in $\mathcal{H}_A$ when the variables A and B are incompatible? Because the condition (24) for orthogonality can be satisfied only for certain very restricted values of the probability functions: specifically, we must have $f_{a_+}\left(\theta_{b_+a_+}\right) = f_{a_-}\left(\theta_{b_-a_-}\right)$, which is only possible when certain symmetries are present. And why is non-orthogonality not an issue when the variables A and B are compatible? Because when the variables are compatible, the eigenstates of A are also eigenstates of B, so the $(b_i)$ states and $(a_i)$ states are one and the same, namely $(a_{\pm}, b_{\pm})$ or $(a_{\pm}, b_{\mp})$.

Returning to equations (23) and introducing the $C_2$ symmetry of the system (the Arrow itself) and of the apparatus (the magnets in specific directions that are used to make measurements of the state of the arrow) results in the following symmetries between the *angles* (Figure 3),

$$\theta_{b_+a_+} = \theta_{b_-a_-} \tag{25a}$$

$$\theta_{a_-b_+} = \theta_{a_+b_-} = \pi - \theta \tag{25b}$$

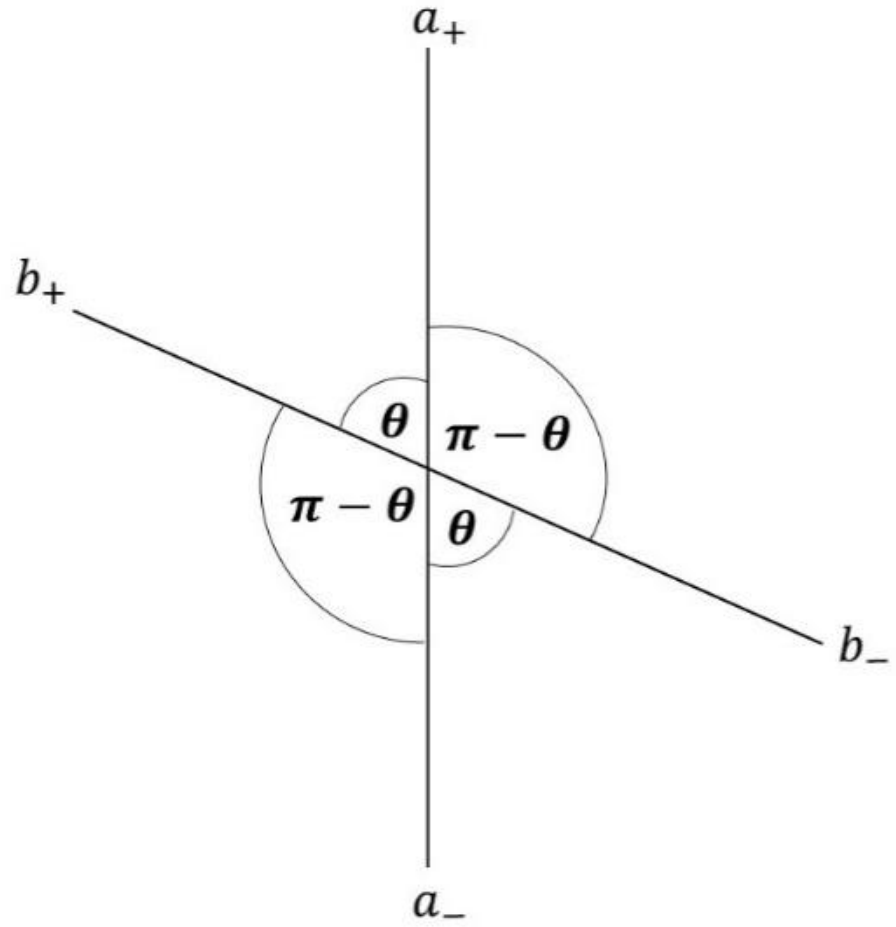

**Figure 3.** Symmetries between angles as a result of $C_2$ apparatus symmetry in the Arrow system

so that equations (23) now become

$$|b_+^A\rangle = f_{a_+}^{1/2}(\theta)|a_+^A\rangle + \left(1 - f_{a_+}(\theta)\right)^{1/2} |a_-^A\rangle \tag{26a}$$

$$|b_-^A\rangle = -\left(1 - f_{a_-}(\theta)\right)^{1/2} |a_+^A\rangle + f_{a_-}^{1/2}(\theta)|a_-^A\rangle \tag{26b}$$

where $\theta$ is the angle between the $\vec{\boldsymbol{a}}$ and $\vec{\boldsymbol{b}}$ directions in physical space. However, $|b_+^A\rangle$ and $|b_-^A\rangle$ are still not orthogonal. Although the *angles* show these symmetries, the *probabilities* do not necessarily show the same symmetries, e.g.

$P_{(b_+)}(a_+) = f_{a_+}(\theta)$ might be different from $P_{(b_-)}(a_-) = f_{a_-}(\theta)$ if there was an apparatus asymmetry between the $a_+$ and $a_-$ directions. An example of an apparatus asymmetry would be having the magnets in the $a_+$ direction stronger than those in the $a_-$ direction, resulting in the probability functions $f_{a_+}$ and $f_{a_-}$ being unequal. But even *with* a $C_2$ *apparatus* symmetry between the $a_+$ and $a_-$ directions, the probability functions $f_{a_+}$and $f_{a_-}$ still cannot be equated unless the $C_2$ symmetry is extended to the *spacetime* that includes the two directions, giving

$$f_{a_+}(\theta) = f_{a_-}(\theta) \equiv f_A(\theta) \tag{27}$$

which results in equations (26) becoming

$$|b_+^A\rangle = f_A^{1/2}(\theta)|a_+^A\rangle + \left(1 - f_A(\theta)\right)^{1/2}|a_-^A\rangle \tag{28a}$$

$$|b_-^A\rangle = -\left(1 - f_A(\theta)\right)^{1/2}|a_+^A\rangle + f_A^{1/2}(\theta)|a_-^A\rangle \tag{28b}$$

We see that the vectors representing the eigenstates $(b_+)$ and $(b_-)$ of B are now orthogonal not only in Hilbert space $\mathcal{H}_B$ (since $\langle b_-^B|b_+^B\rangle = 0$) but also in Hilbert space $\mathcal{H}_A$ (we now have $\langle b_-^A|b_+^A\rangle = 0$, as in equation (24), but *only* because some degree of *symmetry* has been introduced!)

To see this diagrammatically, we note that in the absence of any symmetry, the basis $\{|b_+^A\rangle, |b_-^A\rangle\}$ in $\mathcal{H}_A$ is rotated with respect to the basis $\{|a_+^A\rangle, |a_-^A\rangle\}$ by a *non-rigid* rotation, in which $|b_+^A\rangle$ and $|b_-^A\rangle$ are rotated through *different* angles, making them *non-orthogonal* in $\mathcal{H}_A$ (Figure 4).

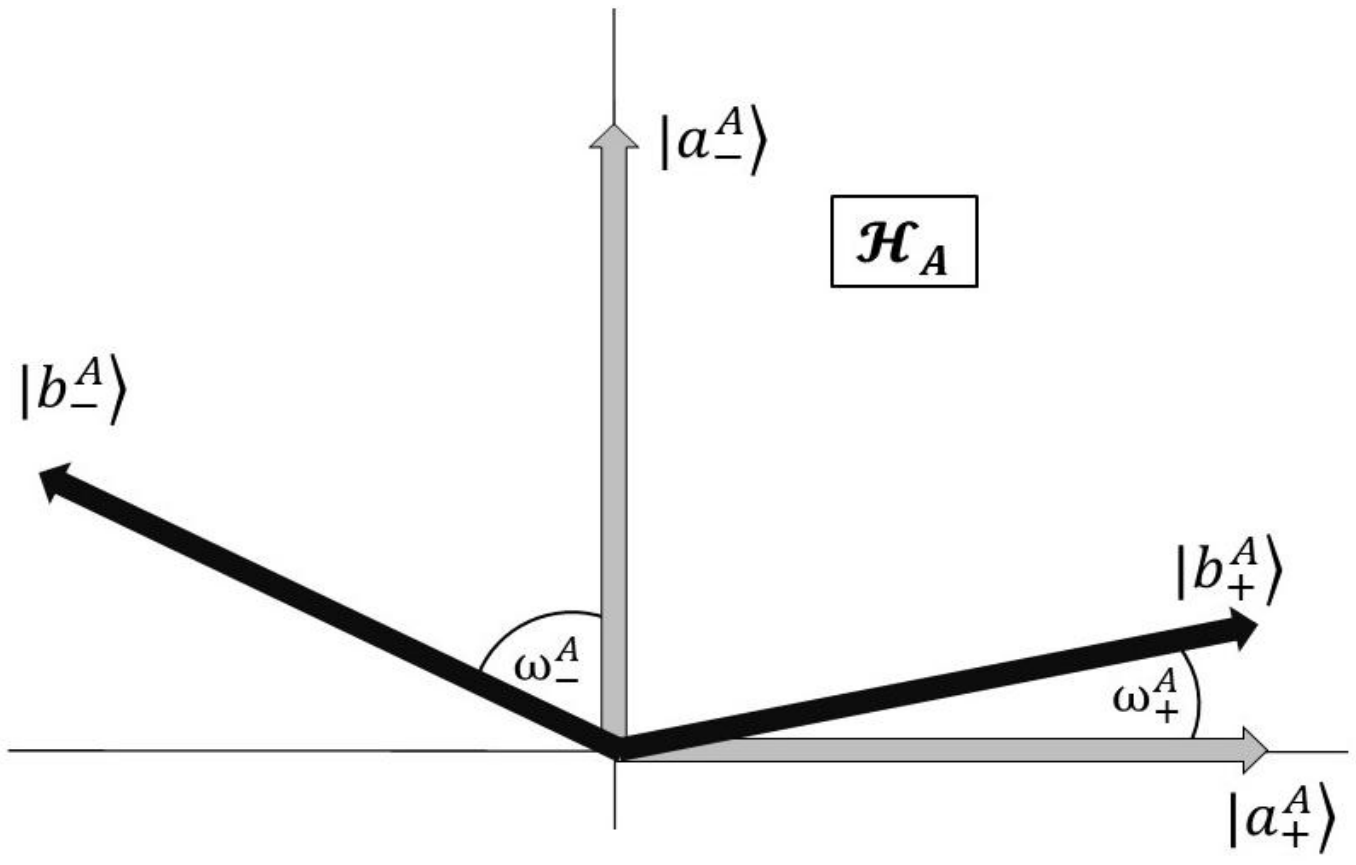

**Figure 4.** In the absence of symmetry, the basis $\{|b_+^A\rangle, |b_-^A\rangle\}$ in the Arrow system is in general rotated with respect to the basis $\{|a_+^A\rangle, |a_-^A\rangle\}$ by a *non-rigid* rotation in Hilbert space $\mathcal{H}_A$.

The two basis sets can be written in column vector form as

$$\boldsymbol{a}_+^A = \begin{pmatrix} 1 \\ 0 \end{pmatrix}, \ \boldsymbol{a}_-^A = \begin{pmatrix} 0 \\ 1 \end{pmatrix} \tag{29}$$

and

$$\boldsymbol{b}_{+}^{A} = \begin{pmatrix} f_{a_+}^{1/2}(\theta) \\ \left(1 - f_{a_+}(\theta)\right)^{1/2} \end{pmatrix}, \;\; \boldsymbol{b}_{-}^{A} = \begin{pmatrix} -\left(1 - f_{a_-}(\theta)\right)^{1/2} \\ f_{a_-}^{1/2}(\theta) \end{pmatrix} \tag{30}$$

and are connected in Hilbert space $\mathcal{H}_A$ by

$$\boldsymbol{b}_{\pm}^{A} = \mathbf{M}_{\mathbf{ab}}^{\mathrm{A}} \boldsymbol{a}_{\pm}^{A} \tag{31}$$

where the rotation matrix $\mathbf{M}_{\mathbf{ab}}^{\mathrm{A}}$ has the form

$$\mathbf{M}_{\mathbf{ab}}^{\mathrm{A}} = \begin{pmatrix} \cos\omega_{+}^{A} & -\sin\omega_{-}^{A} \\ \sin\omega_{+}^{A} & \cos\omega_{-}^{A} \end{pmatrix} \tag{32}$$

with the respective rotation angles given by

$$\omega_{+}^{A} = \tan^{-1}\frac{\left(1 - f_{a_+}(\theta)\right)^{1/2}}{f_{a_+}^{1/2}(\theta)} \tag{33a}$$

$$\omega_{-}^{A} = \tan^{-1}\frac{\left(1 - f_{a_-}(\theta)\right)^{1/2}}{f_{a_-}^{1/2}(\theta)} \tag{33b}$$

Introduction of $C_2$ symmetry of spacetime enabled $f_{a_+}$ and $f_{a_-}$ to be equated,

$$f_{a_+}(\theta) = f_{a_-}(\theta) \equiv f_A(\theta) \tag{34}$$

so that $|b_{+}^{A}\rangle$ and $|b_{-}^{A}\rangle$ are now obtained by rotation of $|a_{+}^{A}\rangle$ and $|a_{-}^{A}\rangle$ through the *same* angle

$$\omega_{+}^{A} = \omega_{-}^{A} = \tan^{-1}\frac{\left(1 - f_{A}(\theta)\right)^{1/2}}{f_{A}^{1/2}(\theta)} = \omega^{A} \tag{35}$$

making the vectors representing $(b_+)$ and $(b_-)$ orthogonal not only in $\mathcal{H}_B$ but also in $\mathcal{H}_A$ (Figure 5).

A key point that will be very important later is that, when *symmetry* is introduced, the rotation matrix in Hilbert space $\mathcal{H}_A$ becomes *unitary*,

$$\mathbf{M}_{\mathbf{ab}}^{\mathrm{A}} = \begin{pmatrix} \cos\omega^{A} & -\sin\omega^{A} \\ \sin\omega^{A} & \cos\omega^{A} \end{pmatrix} \tag{36}$$

and multiplication by a *unitary* matrix preserves both lengths and angles, thus effecting a *rigid* rotation. Similarly, if we consider the situation in Hilbert space $\mathcal{H}_B$, we have

$$\boldsymbol{a}_{\pm}^{B} = \mathbf{M}_{\mathbf{ba}}^{\mathrm{B}} \boldsymbol{b}_{\pm}^{B} \tag{37}$$

and when $C_2$ symmetry is introduced the rotation matrix $\mathbf{M}_{\mathbf{ba}}^{\mathrm{B}}$ becomes unitary,

$$\mathbf{M}_{\mathbf{ba}}^{\mathrm{B}} = \begin{pmatrix} \cos\omega^{B} & -\sin\omega^{B} \\ \sin\omega^{B} & \cos\omega^{B} \end{pmatrix} \tag{38}$$

making the vectors representing $(a_+)$ and $(a_-)$ orthogonal not only in $\mathcal{H}_A$ but also in $\mathcal{H}_B$. However, although $C_2$ symmetry of spacetime allows us to equate $f_{a_+}$ with $f_{a_-}$ and $f_{b_+}$ with $f_{b_-}$,

$$f_{a_+}(\theta) = f_{a_-}(\theta) \equiv f_A(\theta) \tag{39a}$$

$$f_{b_+}(\theta) = f_{b_-}(\theta) \equiv f_B(\theta) \tag{39b}$$

making both $\mathbf{M}^{\mathrm{A}}_{\mathbf{ab}}$ and $\mathbf{M}^{\mathrm{B}}_{\mathbf{ba}}$ unitary, it does *not* allow us to equate $f_A(\theta)$ with $f_B(\theta)$.

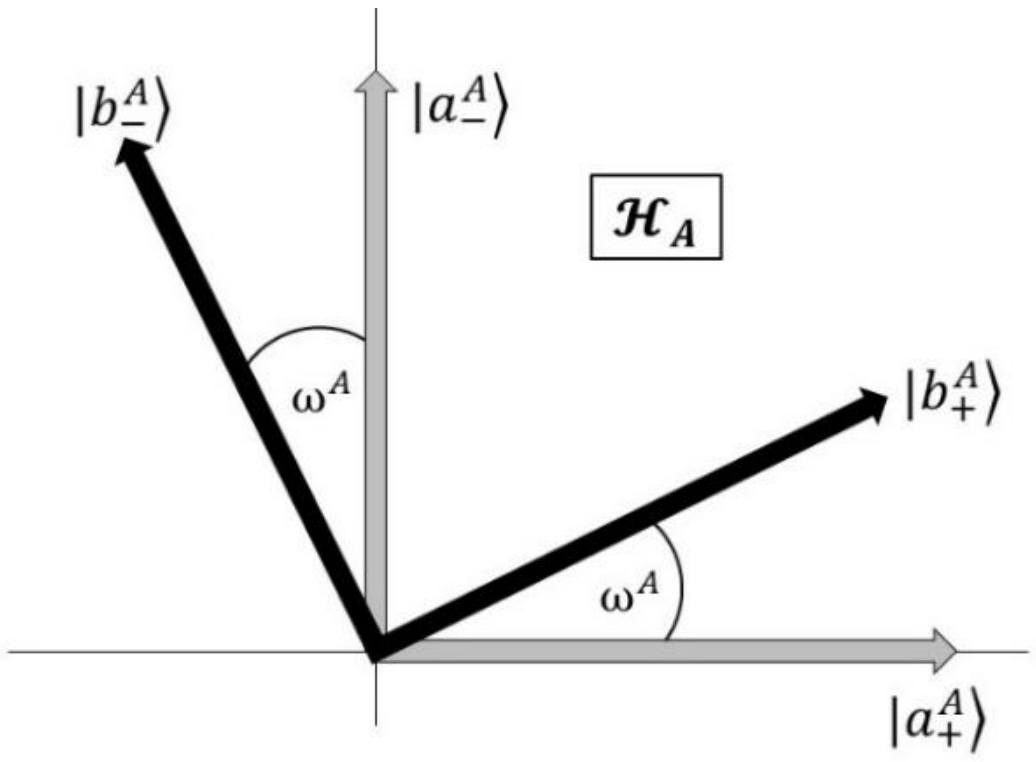

**Figure 5** In the presence of C2 symmetry, the basis $\{|b_+^A\rangle, |b_-^A\rangle\}$ in Hilbert space $\mathcal{H}_A$ is rotated with respect to the basis $\{|a_+^A\rangle, |a_-^A\rangle\}$ by a *rigid* rotation through an angle $\omega^A$.

And we see that if $f_A(\theta) \neq f_B(\theta)$, the angle $\omega^A$ in $\mathcal{H}_A$ (Figure 5) is *not* the same as the angle $\omega^B$ in $\mathcal{H}_B$ (Figure 6),

$$\omega^A = \tan^{-1}\frac{\left(1-f_A(\theta)\right)^{1/2}}{f_A^{1/2}(\theta)} \qquad (40a)$$

$$\omega^B = \tan^{-1}\frac{\left(1-f_B(\theta)\right)^{1/2}}{f_B^{1/2}(\theta)} \qquad (40b)$$

which means that the two Hilbert spaces are *not equivalent*, because they give different results for probabilities, for example $|\langle b_+^A|a_+^A\rangle|^2$ and $|\langle b_+^B|a_+^B\rangle|^2$ respectively, for the probability of getting the result $b_+$ when starting from state $(a_+)$.

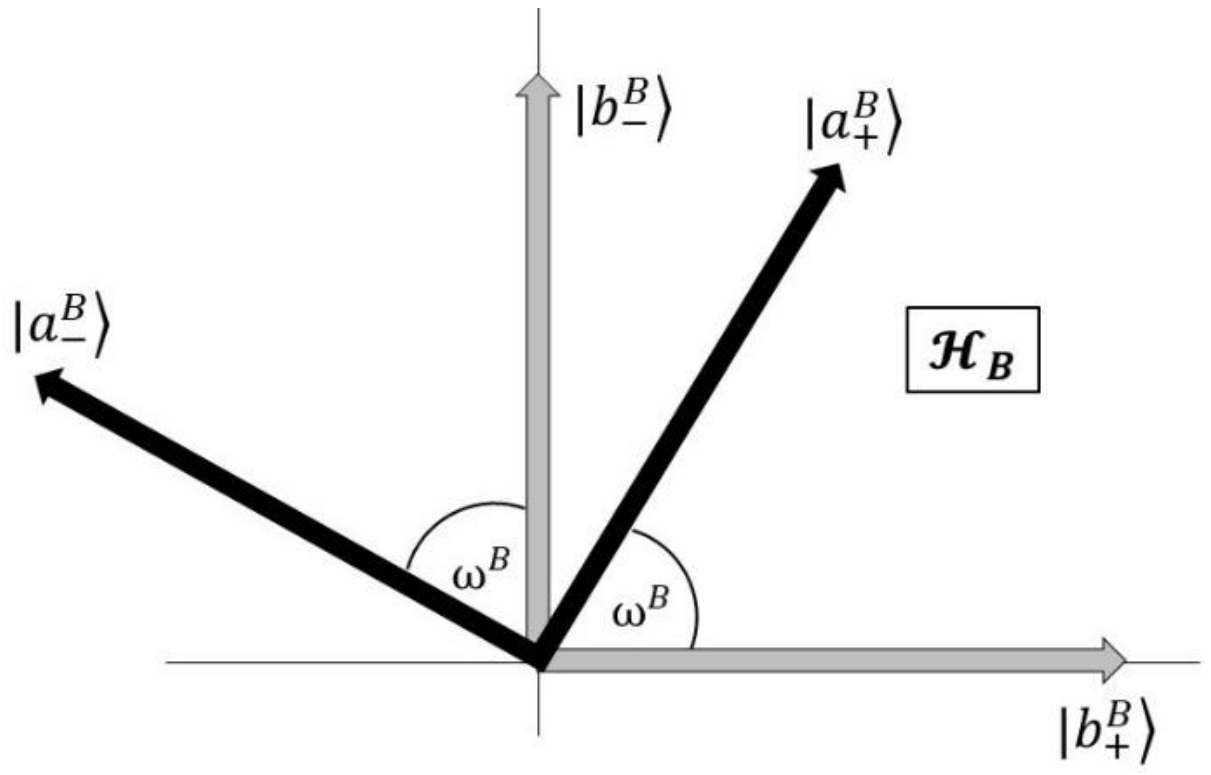

**Figure 6** In the presence of C2 symmetry, the basis $\{|a_+^B\rangle, |a_-^B\rangle\}$ of the Arrow system in Hilbert space $\mathcal{H}_B$ is rotated with respect to the basis $\{|b_+^B\rangle, |b_-^B\rangle\}$ by a rigid rotation through an angle $\omega^B$ that is *not the same* as the corresponding rotation angle $\omega^A$ in Hilbert space $\mathcal{H}_A$

So even though extending the C2 symmetry of the system-plus-apparatus to spacetime using equations (39) *does* make *both* the vectors representing $(a_+)$ and $(a_-)$ *and* the vectors representing $(b_+)$ and $(b_-)$ orthogonal *in both* $\mathcal{H}_A$ *and* $\mathcal{H}_B$, we must still calculate A measurement probabilities only in $\mathcal{H}_A$ and B measurement probabilities only in $\mathcal{H}_B$. In other words the Hilbert spaces $\mathcal{H}_A$ and $\mathcal{H}_B$ are *still* not equivalent with only $C_2$ symmetry because the orthogonal vectors representing $(a_+)$ and $(a_-)$ are related to the orthogonal vectors representing $(b_+)$ and $(b_-)$ by rigid rotations through angles $\omega^A$ and $\omega^B$ respectively that are *not the same* in the two Hilbert spaces (Figures 5 and 6).

To be able to calculate probabilities of both A and B outcomes in either $\mathcal{H}_A$ or $\mathcal{H}_B$, we have to extend the symmetry of spacetime from $C_2$ to full *isotropy* to make every direction A, B, C,… equivalent,

$$f_A(\theta) = f_B(\theta) = f_C(\theta) = \ldots \equiv f(\theta) \tag{41}$$

in order to equate $\omega^A$ with $\omega^B$ to give

$$\omega^A = \omega^B = \omega \tag{42}$$

where

$$\omega = \omega(\theta) = \tan^{-1} \frac{\left(1-f(\theta)\right)^{1/2}}{f^{1/2}(\theta)} \tag{43}$$

Once the vectors representing $(a_+)$ and $(a_-)$ *and* the vectors representing $(b_+)$ and $(b_-)$ become orthogonal *in both* $\mathcal{H}_A$ *and* $\mathcal{H}_B$ (after introduction of $C_2$ symmetry), and also the probability functions $f_A$, $f_B$, $f_C$, … become identical (after introduction of complete isotropy), probabilities of $b_i$ outcomes can then be correctly calculated in either $\mathcal{H}_A$ or $\mathcal{H}_B$ (as can probabilities of $a_i$ outcomes). It is then no longer necessary to have two Hilbert spaces and it becomes possible to combine the Hilbert spaces $\mathcal{H}_A$ and $\mathcal{H}_B$ into one. And if all the states $(a_i)$ and $(b_j)$ *can* now be represented in just one Hilbert space thanks to $C_2$ symmetry of system and apparatus, plus complete isotropy of spacetime, it means that according to Gleason's theorem we *must* now write probabilities in the Born form as $\mathrm{P}_{(b_j)}(a_i) = \left|\langle a_i | b_j \rangle\right|^2 = \cos^2 \omega(\theta)$!

But it is important to distinguish clearly between the *physical* angle $\theta$ between the $b_j$ and $a_i$ directions in ordinary physical space and the *Pythagorean* angle $\omega(\theta)$ between the corresponding vectors $|b_i\rangle$ and $|a_i\rangle$ in Hilbert space: although the angle $\omega$ is a function of $\theta$, the fact that we now have Born "cosine-squared" dependence on $\omega$ does not in any way imply a "cosine-squared" dependence of $f(\theta)$ on $\theta$. However, having $|b_+^A\rangle$ and $|b_-^A\rangle$ orthogonal when $C_2$ symmetry is introduced does place an additional constraint $f(\theta) + f(\pi - \theta) = 1$ on the form of the probability function (see equations (25) and Figure 3) that now excludes $f(\theta) = 1 - \left(\frac{\theta}{\pi}\right)^2$ as a probability function, but still allows $f(\theta) = 1 - \frac{\theta}{\pi}$ and $f(\theta) = \cos^2 ½\theta$. So introducing symmetry into the Arrow system to produce a single Hilbert space with a single unrestricted Born Rule is one thing, but uniquely predicting the form of $f(\theta)$ would seem to be quite another.

However, complete isotropy of space implies more than just the same probability function $f_A(\theta) = f_B(\theta) = f_C(\theta) = \ldots \equiv f(\theta)$ for all directions A, B, C… . It also has the further implication that the results of measurements of any variable A (or B or C…) must be unchanged if the *entire experiment*, i.e. the system (the Arrow along the $\vec{\boldsymbol{n}}$ axis) together with the apparatus (the magnets along the $\vec{\boldsymbol{a}}$ axis), is rotated to a new orientation in space. In other words, we must ensure that physical rotation of the experiment through any angle $\alpha$ about the *y*-axis (i.e. perpendicular to the arrow) is a *symmetry transformation*, and we will show that enforcing this requirement *does* uniquely fix the form of $f(\theta)$.

Symmetry transformations leave a system "looking the same", so they have to preserve lengths and angles in physical space, which means they must be *rigid*; they must also leave experimental results unchanged, which means that the *rigidity* of symmetry transformations in *physical* space must carry over to a similar rigidity in *Hilbert* space, in order to conserve probabilities (which depend on angles in Hilbert space). Wigner proved in his famous theorem that symmetry transformations in Hilbert space are effected by operators that are *linear* and *unitary*. But Wigner's proof [14,17] implicitly assumes a *single* Hilbert space, so his theorem is not (except in special cases) applicable to general incompatible variables, which require *multiple* Hilbert spaces. However, there is another symmetry condition that we expect to be obeyed in every separate Hilbert space of the GIV theory. This is the *Haag-Wigner condition* [17,18] (see Appendix 2), which states that the operators that effect symmetry transformations in a Hilbert space must depend only on the *relation* between the two frames of reference in physical space-time, not on the intrinsic properties or absolute position in space-time of either frame, otherwise the homogeneity and isotropy of space-time would be violated. The simplest way to satisfy this condition is to require the symmetry operator to be *linear*; and a linear symmetry operator in a Hilbert space in which all vectors representing physical states are required to be normalized will automatically also be *unitary* [19, 20], thus ensuring that the transformation effected by the operator is *rigid*, as required of a symmetry transformation.

In the general case (see Section 5) we will satisfy the Haag-Wigner condition for symmetry operators by directly requiring them to be linear (and therefore unitary). In the specific case of the Arrow system we satisfy the Haag-Wigner condition by considering a rotation of the entire experiment (i.e. both Arrow $\vec{\boldsymbol{n}}$ and apparatus $\vec{\boldsymbol{a}}$) through an angle $\alpha$, and requiring the matrix $\mathbf{Y}^{\mathrm{A}}$ that effects this rotation to be linear, i.e. dependent only on the angle $\alpha$ of rotation in physical space, and independent of the different starting angles of Arrow and apparatus. This linearity requirement places further restrictions on the form of the probability function $f(\theta) = \cos^2 \omega(\theta)$, where $\omega(\theta)$ is the Pythagorean angle in Hilbert space $\mathcal{H}_A$ corresponding to a physical angle $\theta$ between $\vec{\boldsymbol{n}}$ and $\vec{\boldsymbol{a}}$ in physical space, and we show in Appendix 3 that these restrictions yield exclusively the simple form

$$\omega(\theta) = ½\theta \tag{44}$$

and also the equivalence of all of the Hilbert spaces $\mathcal{H}_A, \mathcal{H}_B, \ \mathcal{H}_C \dots$ . Complete isotropy of space thus restricts the probability function to the unique form

$$f(\theta) = \cos^2 \omega(\theta) = \ \cos^2 ½\theta \tag{45}$$

and requires only a single Hilbert space, which is precisely the result for the quantum mechanics of a spin ½ system – and yet this result emerges for an entirely *classical* probabilistic system possessing incompatible variables (our Arrow) when we apply certain particular symmetry constraints! Hence we have the situation that in principle we can construct a classical Arrow system with almost any choice for the probability function $f(\theta)$ (and with most choices not having the symmetries we have discussed), whereas a spin ½ particle apparently has no choice but to incorporate the symmetries of spacetime.

# 5 The General Incompatible Variables (GIV) theory

The Arrow model introduced in Section 4 provided a means to uncover and highlight the role of symmetry in probabilistic theories of systems possessing incompatible variables. We now generalize our treatment of the classical Arrow system to provide a general framework – the General Incompatible Variables (GIV) formalism – for the treatment of any system, whether QM or non-QM, that has incompatible variables. We will see that QM is just a special case of GIV theories in which the *fundamental variables* are *symmetries*. The GIV formalism is basically the standard classical probability theory of Kolmogorov extended to include incompatible variables with the introduction of a separate Hilbert space for each incompatible variable. We first demonstrate the GIV formalism for the simple case of just two incompatible variables, A and B, each of which has just two possible measurement outcomes, $a_1$ and $a_2$ for A, and $b_1$ and $b_2$ for B. Each state $(\psi)$ of the system can be thought of as having "components" in each of the separate Hilbert spaces:

$$(\psi) = (|\psi^A\rangle, |\psi^B\rangle) \tag{46}$$

The fundamental pure states (also called value states or eigenstates) of this system are then given in the GIV formalism as

$$(a_i) = \left(|a_i^A\rangle, |a_i^B\rangle\right), \ \ i = 1, 2 \tag{47a}$$

$$(b_i) = (|b_i^A\rangle, |b_i^B\rangle), \ \ \ i = 1, 2 \tag{47b}$$

which requires two Hilbert spaces, $\mathcal{H}_A$ and $\mathcal{H}_B$, where $\mathcal{H}_A$ is spanned by the orthonormal basis $\{|a_1^A\rangle, |a_2^A\rangle\}$ and $\mathcal{H}_B$ by the orthonormal basis $\{|b_1^B\rangle,|b_2^B\rangle\}$.

Note that the vectors representing the states $(b_1)$ and $(b_2)$ are orthogonal in $\mathcal{H}_B$, but are (generally) *not* orthogonal in $\mathcal{H}_A$ (as we saw in Figure 4 for the Arrow), so the probabilities of the outcomes $b_1$ and $b_2$ cannot in general be calculated in $\mathcal{H}_A$, they can only be calculated in $\mathcal{H}_B$. Similarly, the vectors representing the states $(a_1)$ and $(a_2)$ are orthogonal in $\mathcal{H}_A$, but are (generally) *not* orthogonal in $\mathcal{H}_B$, so the probabilities of the outcomes $a_1$ and $a_2$ can only be calculated in $\mathcal{H}_A$. Choosing corresponding axes of $\mathcal{H}_A$ and $\mathcal{H}_B$ to be parallel (or superimposed), we write the basis vectors of $\mathcal{H}_A$ as

$$\boldsymbol{a}_1^A = \begin{pmatrix}1\\0\end{pmatrix},\ \boldsymbol{a}_2^A = \begin{pmatrix}0\\1\end{pmatrix}, \tag{48a}$$

the basis vectors of $\mathcal{H}_B$ as

$$\boldsymbol{b}_1^B = \begin{pmatrix}1\\0\end{pmatrix},\ \boldsymbol{b}_2^B = \begin{pmatrix}0\\1\end{pmatrix} \tag{48b}$$

and the variables A and B as the matrices

$$\mathbf{A_A} = \begin{pmatrix}a_1 & 0\\0 & a_2\end{pmatrix} \tag{49a}$$

$$\mathbf{B_B} = \begin{pmatrix}b_1 & 0\\0 & b_2\end{pmatrix} \tag{49b}$$

where the subscripts A and B indicate that the matrix $\mathbf{A}_\mathrm{A}$ acts as an operator in $\mathcal{H}_A$, and the matrix $\mathbf{B}_\mathrm{B}$ acts as an operator in $\mathcal{H}_B$. We then have the eigenvalue equations

$$\mathbf{A_A}\boldsymbol{a}_i^A = a_i\boldsymbol{a}_i^A\ ,\ i = 1,2 \tag{50a}$$

$$\mathbf{B_B}\boldsymbol{b}_i^B = b_i\boldsymbol{b}_i^B\ ,\ i = 1,2 \tag{50b}$$

in matrix notation, and

$$\hat{A}_A|a_i^A\rangle = a_i|a_i^A\rangle,\ i = 1,2 \tag{51a}$$

$$\hat{B}_B|b_i^B\rangle = b_i|b_i^B\rangle,\ i = 1,2 \tag{51b}$$

in Dirac notation. This notation is extended in an obvious way when there are additional incompatible variables C, D,… and more than two possible values for the variables. Finally, we note that equations (48) exhibit a simple feature that will become crucial later on, namely that

$$\boldsymbol{a}_1^A = \boldsymbol{b}_1^B = \begin{pmatrix}1\\0\end{pmatrix} \tag{52a}$$

$$\boldsymbol{a}_2^A = \boldsymbol{b}_2^B = \begin{pmatrix}0\\1\end{pmatrix} \tag{52b}$$

and, more generally,

$$\boldsymbol{a}_i^A = \boldsymbol{b}_i^B = \boldsymbol{c}_i^C = \cdots,\ i = 1,2,3\ ... \tag{53}$$

We would like to point out that although the use of multiple Hilbert spaces, each with a separate Born rule, may seem an unnecessary complication, it is actually an almost trivial construction that can be applied to any probabilistic theory – and it is essential in the case of incompatible variables, because calculating the outcomes of measurements of B in Hilbert space $\mathcal{H}_A$ and of A in $\mathcal{H}_B$ cannot be done correctly in the general case, as we have seen. Introducing the multiple spaces also provides the most direct route to quantum mechanics as a special case, as we will show.

We now consider how symmetry transformations work in the multiple Hilbert spaces of a GIV system with incompatible variables A, B, C…. if the system has the symmetry of a group

$$\mathrm{G} = \{\mathrm{E}, \mathrm{X}, \mathrm{Y}, \mathrm{Z} \dots\} \tag{54}$$

of symmetries E, X, Y, Z…. where E is the identity element of the group. Applying the symmetry transformation X to the system in any admissible pure state, say $(a_i)$, will transform the state into another admissible pure state [18], say $(a_i')$:

$$\mathrm{X}(a_i) = (a_i') \tag{55}$$

Since the GIV states are sequences of vectors in the Hilbert spaces $\mathcal{H}_A$, $\mathcal{H}_B$, $\mathcal{H}_C$…., the effect of X is to transform the sequence of vectors

$$(a_i) = \left(|a_i^A\rangle, |a_i^B\rangle, |a_i^C\rangle \dots\dots\right) \tag{56}$$

into the sequence

$$(a_i') = \left(|a_i'^A\rangle, |a_i'^B\rangle, |a_i'^C\rangle \dots\dots\right) \tag{57}$$

So equation (55) in fact represents a sequence of equations written in terms of operators acting in the individual Hilbert spaces as

$$\hat{X}_A|a_i^A\rangle = |a_i'^A\rangle \tag{58a}$$

$$\hat{X}_B|a_i^B\rangle = |a_i'^B\rangle \tag{58b}$$

$$\hat{X}_C|a_i^C\rangle = |a_i'^C\rangle \tag{58c}$$

$$\dots \qquad \dots$$

where $\hat{X}_A$ is the operator corresponding to the symmetry X acting in $\mathcal{H}_A$, $\hat{X}_B$ is the corresponding operator in $\mathcal{H}_B$, etc., and similar equations are obtained for any of the symmetries E, X, Y, Z….. in the group G. Since the vector $|a_i'^A\rangle$ is a vector in $\mathcal{H}_A$, it can be expanded in terms of the orthonormal basis $\{|a_k^A\rangle\}$ of $\mathcal{H}_A$, so that equation (58a) becomes

$$\hat{X}_A|a_i^A\rangle = \sum_k |a_k^A\rangle\langle a_k^A|\hat{X}_A|a_i^A\rangle = \sum_k [\Gamma_\mathrm{A}(\mathrm{X})]_{ki}\,|a_k^A\rangle \tag{59}$$

where the expansion coefficients are the elements

$$[\Gamma_\mathrm{A}(\mathrm{X})]_{ki} = \langle a_k^A|\hat{X}_A|a_i^A\rangle \tag{60}$$

of a square matrix $\mathbf{\Gamma_A}(\mathrm{X})$. Similar matrices $\mathbf{\Gamma_A}(\mathrm{E})$, $\mathbf{\Gamma_A}(\mathrm{Y})$….. can be defined for all the symmetry elements, and the set of matrices

$$\Gamma_A = \{\, \mathbf{\Gamma_A}(\mathrm{E}), \mathbf{\Gamma_A}(\mathrm{X}), \mathbf{\Gamma_A}(\mathrm{Y}), \mathbf{\Gamma_A}(\mathrm{Z}) \dots\} \tag{61}$$

together form a representation of the group G (using the eigenstates of $\hat{A}$ as basis vectors), provided they satisfy the group multiplication table – but this is only possible if the operators $\hat{X}_A, \hat{Y}_A, \hat{Z}_A$….. are *linear* operators. For example, to obtain the matrix representing YX in the group, we would have, from equation (59),

$$\hat{Y}_A\hat{X}_A|a_i^A\rangle = \hat{Y}_A \sum_k [\Gamma_A(X)]_{ki}\,|a_k^A\rangle \quad (62)$$

and only if $\hat{Y}_A$ is a *linear* operator (i.e. one that preserves the operations of scalar multiplication and addition, see Appendix 2) can we obtain

$$\hat{Y}_A\hat{X}_A|a_i^A\rangle = \sum_k [\Gamma_A(X)]_{ki}\,\hat{Y}_A|a_k^A\rangle \quad (63)$$

which then becomes

$$\hat{Y}_A\hat{X}_A|a_i^A\rangle = \sum_k [\Gamma_A(Y)\Gamma_A(X)]_{ki}\,|a_k^A\rangle \quad (64)$$

showing that the matrix product $\Gamma_A(Y)\Gamma_A(X)$ represents the group element $\hat{Y}_A\hat{X}_A$.

We would like to be able to require linearity of the symmetry operators $\hat{X}_A, \hat{Y}_A, \hat{Z}_A \ldots$ operating in Hilbert space $\mathcal{H}_A$ (and similarly for the symmetry operators $\hat{X}_B, \hat{Y}_B, \hat{Z}_B \ldots$ operating in Hilbert space $\mathcal{H}_B$, and for the corresponding symmetry operators in all the other multiple Hilbert spaces $\mathcal{H}_C$, $\mathcal{H}_D \ldots$.) to ensure that the set of matrices in equation (61) will indeed be a representation of the group G. However, we cannot use Wigner's theorem here to assign linearity and unitarity to all the symmetry operators in the multiple Hilbert spaces of GIV theory, because the proof of Wigner's theorem [14] *assumes* a single Hilbert space – along with its unrestricted Born expression for probability – which is the very thing we wish to *prove*! We instead follow the procedure used in Section 2 for the Arrow, and again appeal to the *Haag-Wigner condition* [17,18], which requires symmetry operators to depend only on the *relation* between two frames of reference, and not on the intrinsic properties of either; this condition is most easily satisfied by operators that are *linear*, and linear operators possessing an inverse and operating in a Hilbert space with vectors normalized to unit magnitude are automatically *unitary* [19,20]. The Haag-Wigner condition thus leads us to require linearity (and hence also unitarity) for the symmetry operators $\hat{X}_A, \hat{Y}_A, \hat{Z}_A \ldots$ operating in Hilbert space $\mathcal{H}_A$ (and similarly for the symmetry operators $\hat{X}_B, \hat{Y}_B, \hat{Z}_B \ldots$ operating in Hilbert space $\mathcal{H}_B$, and similarly for the symmetry operators in all the other Hilbert spaces), and then to investigate where this requirement leads – and we will find that it leads to QM.

The linearity-unitarity of the matrices $\mathbf{\Gamma_A}(\mathrm{X}), \mathbf{\Gamma_A}(\mathrm{Y}), \mathbf{\Gamma_A}(\mathrm{Z}) \ldots$ now confirms that they do indeed represent the operators $\hat{X}_A, \hat{Y}_A, \hat{Z}_A \ldots$ (up to phase factors), so we now have the following sets of linear-unitary representatives of the group $\mathrm{G} = \{\mathrm{E}, \mathrm{X}, \mathrm{Y}, \mathrm{Z} \ldots\}$:

$$\Gamma_A = \{\mathbf{E_A}, \mathbf{X_A}, \mathbf{Y_A}, \mathbf{Z_A} \ldots\} \quad (65a)$$
$$\Gamma_B = \{\mathbf{E_B}, \mathbf{X_B}, \mathbf{Y_B}, \mathbf{Z_B} \ldots\} \quad (65b)$$
$$\Gamma_C = \{\mathbf{E_C}, \mathbf{X_C}, \mathbf{Y_C}, \mathbf{Z_C} \ldots\} \quad (65c)$$
$$\ldots \qquad \ldots$$

where the notation has been simplified by writing the matrices as $\mathbf{X_A}, \mathbf{Y_A} \ldots$ rather than $\mathbf{\Gamma_A}(\mathrm{X}), \mathbf{\Gamma_A}(\mathrm{Y}) \ldots$. The respective representations $\Gamma_A, \Gamma_B, \Gamma_C \ldots$ use the eigenstates of the respective fundamental variables A, B, C… as basis vectors.

We next investigate what happens in the GIV framework when we take the fundamental variables A, B, C… to be the symmetries X, Y, Z… of the system, so that the group $\mathrm{G} = \{\mathrm{E}, \mathrm{X}, \mathrm{Y}, \mathrm{Z} \dots\}$ now becomes

$$\mathrm{G} = \{\mathrm{E}, \mathrm{A}, \mathrm{B}, \mathrm{C} \dots\} \tag{66}$$

The fundamental pure states $(a_i), (b_i), (c_i)$… are the eigenstates of the respective variables A, B, C…, which are now symmetries – and as eigenstates (i.e. "value states") of symmetry operators, they are now states of definite symmetry. So whereas previously a symmetry X changed a general GIV state $(a_i)$ into a different state $(a_i')$ as in equation (55), $\mathrm{X}(a_i) = (a_i')$, the fundamental pure states $(a_i)$ are now *invariant* under the symmetry A, so equation (55) now becomes

$$\mathrm{A}(a_i) = (a_i) \tag{67}$$

and equations (58) now become the set of eigenvalue equations

$$\hat{A}_A |a_i^A\rangle = a_i^A |a_i^A\rangle \tag{68a}$$

$$\hat{A}_B |a_i^B\rangle = a_i^B |a_i^B\rangle \tag{68b}$$

$$\hat{A}_C |a_i^C\rangle = a_i^C |a_i^C\rangle \tag{68c}$$

$$\dots \qquad \dots$$

in their respective Hilbert spaces, with generally complex eigenvalues of unit magnitude since the symmetry operators are unitary, and similarly $\mathrm{B}(b_i) = (b_i)$ represents the set of equations

$$\hat{B}_A |b_i^A\rangle = b_i^A |b_i^A\rangle \tag{69a}$$

$$\hat{B}_B |b_i^B\rangle = b_i^B |b_i^B\rangle \tag{69b}$$

$$\hat{B}_C |b_i^C\rangle = b_i^C |b_i^C\rangle \tag{69c}$$

$$\dots \qquad \dots$$

For Lie groups G [14] the operators $\hat{A}_A, \hat{A}_B \dots, \hat{B}_A, \hat{B}_B \dots$ can be taken to be either the symmetry operators themselves or their generators, because the eigenstates of the symmetry operators are also eigenstates of the generators, e.g. in the case of rotational symmetry in QM, the angular momentum operator $\hat{J}_z$ is the generator of rotations $\hat{R}^z(\phi) = e^{-(i/\hbar)\phi \hat{J}_z}$ through an angle $\phi$ about the *z*-axis, and the angular momentum states $|J, M\rangle$ are eigenstates of both $\hat{R}^z(\phi)$ and $\hat{J}_z$.

Applying the linearity-unitarity requirement to all the symmetry operators $\hat{A}_A, \hat{A}_B \dots,\ \hat{B}_A, \hat{B}_B \dots$ in each of the multiple Hilbert spaces now gives us the following sets of linear-unitary representatives of the group $\mathrm{G} = \{\mathrm{E}, \mathrm{A}, \mathrm{B}, \mathrm{C} \dots\}$:

$$\Gamma_A = \{\mathbf{E_A}, \mathbf{A_A}, \mathbf{B_A}, \mathbf{C_A} \dots\} \tag{70a}$$

$$\Gamma_B = \{\mathbf{E_B}, \mathbf{A_B}, \mathbf{B_B}, \mathbf{C_B} \dots\} \tag{70b}$$

$$\Gamma_C = \{\mathbf{E_C}, \mathbf{A_C}, \mathbf{B_C}, \mathbf{C_C} \dots\} \tag{70c}$$

$$\dots \qquad \dots$$

which are equations (65) with the general symmetries X, Y, Z… replaced by the symmetries A, B, C… which are now also the variables.

We focus here on equations (70) as they apply to *irreducible representations*, which we hereafter refer to as "irreps" [21], of the group G. We will see that these irreps describe *elementary systems*, in line with Wigner's discovery [14,18] that the manifold of states of an *elementary* particle constitute a representation space for an irrep of the appropriate group of symmetries of nature (the Poincaré group for relativistic elementary particles, the full rotation group for electrons in one-electron atoms, and so on). A familiar example is provided by the hydrogenic s, p, d… electrons: their angular wavefunctions are the spherical harmonics, which form the basis functions for the respective irreps $\Gamma^{(0)}$, $\Gamma^{(1)}$, $\Gamma^{(2)}$ ... of the full rotation group $\mathrm{R}_3$. In standard QM there is only one Hilbert space, and only one irrep, for a given elementary physical system (e.g. only one irrep $\Gamma^{(0)}$ for an s-electron). But in GIV theory, there are many Hilbert spaces $\mathcal{H}_A$, $\mathcal{H}_B, \mathcal{H}_C \dots$, and equally many irreps $\Gamma_A, \Gamma_B, \Gamma_C \dots$, all applying to the same physical system, each based on a different set of basis vectors (the eigenvectors of the incompatible variables A, B, C… respectively). So the representations in equations (70) are now taken to be the irreps $\Gamma_A, \Gamma_B, \Gamma_C$ … based on the sets of basis vectors $\{|a_i^A\rangle\}$, $\{|b_i^B\rangle\}$, $\{|c_i^C\rangle\}\dots$ of the Hilbert spaces $\mathcal{H}_A$, $\mathcal{H}_B, \mathcal{H}_C \dots$ respectively.

To make further progress and to be consistent with well-known experimental facts, we now require these irreps $\Gamma_A, \Gamma_B, \Gamma_C$ … in the different Hilbert spaces to be *equivalent*. The existence of "equivalent" matrix representations describing the effect of symmetry operations on alternate sets of basis functions is not unusual in standard QM, and arises naturally when variables are incompatible. For example, the basis vectors for irrep $\Gamma^{(J)}$ of the full rotation group $\mathrm{R}_3$ are the vectors $|J, M\rangle$, which are normally taken to be the eigenstates $|J, M_z\rangle$ of $\hat{J}^2$ and $\hat{J}_z$. But the eigenstates $|J, M_x\rangle$ of $\hat{J}_x$, which does not commute with $\hat{J}_z$, would form equally valid or "equivalent" basis vectors for the irrep $\Gamma^{(J)}$, as would the eigenstates $|J, M_y\rangle$ of $\hat{J}_y$, leading to an identical character table and thence identical physical results. Another example of the eigenstates of incompatible variables forming alternate but "equivalent" basis functions for irreps occurs in the Poincaré group, where the eigenvectors $|k\rangle$ of the spatial translation operator are normally taken to be the basis vectors for the irreps, but one could alternatively (and equivalently) use the eigenvectors $|K\rangle$ of the Lorentz boost operator, which does not commute with the spatial translation operator. In standard QM, the definition of the *equivalence* of two representations $\Gamma_\varphi$ and $\Gamma_{\varphi'}$, based respectively on two different sets of basis vectors $\{\boldsymbol{\varphi}_i\}$ and $\{\boldsymbol{\varphi}'_i\}$, is that corresponding matrix representatives $\boldsymbol{\Gamma}_{\boldsymbol{\varphi}}(\mathrm{X})$ and $\boldsymbol{\Gamma}_{\boldsymbol{\varphi}'}(\mathrm{X})$ for a given symmetry operation X of the group are related by a *similarity transformation*

$$\boldsymbol{\Gamma}_{\boldsymbol{\varphi}'}(\mathrm{X}) = \mathbf{S}\,\boldsymbol{\Gamma}_{\boldsymbol{\varphi}}(\mathrm{X})\,\mathbf{S}^{-1} \tag{71}$$

or, in simplified notation,

$$\mathbf{X}' = \mathbf{S}\,\mathbf{X}\,\mathbf{S}^{-1} \tag{72}$$

where $\mathbf{S}$ is the matrix that relates individual basis vectors $\boldsymbol{\varphi}_i$ of $\Gamma_\varphi$ to the corresponding individual basis vectors $\boldsymbol{\varphi}'_i$ of $\Gamma_{\varphi'}$

$$\boldsymbol{\varphi}'_i = \mathbf{S}\boldsymbol{\varphi}_i\ ,\ \ i = 1, 2, 3 \ldots \tag{73}$$

and $\boldsymbol{\varphi}_i$ and $\boldsymbol{\varphi}'_i$ are column vectors in the single Hilbert space of standard QM.

So we require that the irreps $\Gamma_A, \Gamma_B, \Gamma_C \ldots$ in equations (70), based respectively on the eigenvectors of the incompatible variables A, B, C…, all be equivalent to one other, where our *generalized* definition of equivalence in GIV theory is as follows: the irrep $\Gamma_A$ in Hilbert space $\mathcal{H}_A$ is equivalent to the corresponding irrep $\Gamma_B$ in Hilbert space $\mathcal{H}_B$ if (1) the dimensions $n_A$ of $\mathcal{H}_A$ and $n_B$ of $\mathcal{H}_B$ are equal, $n_A = n_B = n$, and (2) the irreps are related by a generalized similarity transformation

$$\boldsymbol{\Gamma}_\mathbf{B} = \mathbf{S}_\mathbf{AB}\boldsymbol{\Gamma}_\mathbf{A}\mathbf{S}_\mathbf{AB}^{-1} \tag{74}$$

that relates every matrix representative $\boldsymbol{\Gamma}_\mathbf{A}$ in the irrep $\Gamma_A$ to the corresponding matrix representative $\boldsymbol{\Gamma}_\mathbf{B}$ in the irrep $\Gamma_B$. We call equation (74) a *generalized* similarity transformation because it is analogous to the regular similarity transformation relation (71) that defines equivalence in QM, except that the matrix $\mathbf{S}_\mathbf{AB}$ in equation (74) does not operate within a single Hilbert space, as in QM, but instead maps matrices $\boldsymbol{\Gamma}_\mathbf{A}$ that act in Hilbert space $\mathcal{H}_A$ onto the corresponding matrices $\boldsymbol{\Gamma}_\mathbf{B}$ that act in Hilbert space $\mathcal{H}_B$. In Appendix 4 we justify this *generalized equivalence requirement* based on requiring equal Casimir invariants for $\Gamma_A$ and $\Gamma_B$ to avoid violating well-established experimental observations, such as the fact that an elementary particle has a fixed value of its spin J, independent of the spatial direction of its spin quantization axis, e.g. an elementary particle cannot be a spin zero particle for measurements in the $z$ direction and a spin one particle for measurements in the $x$ direction!

Equation (74) relates every matrix representative $\boldsymbol{\Gamma}_\mathbf{A}$ in the irrep $\Gamma_A$ to the corresponding matrix representative $\boldsymbol{\Gamma}_\mathbf{B}$ in the irrep $\Gamma_B$ by the mappings

$$\mathbf{A}_\mathbf{B} = \mathbf{S}_\mathbf{AB}\mathbf{A}_\mathbf{A}\mathbf{S}_\mathbf{AB}^{-1} \tag{75a}$$

$$\mathbf{B}_\mathbf{B} = \mathbf{S}_\mathbf{AB}\mathbf{B}_\mathbf{A}\mathbf{S}_\mathbf{AB}^{-1} \tag{75b}$$

$$\mathbf{C}_\mathbf{B} = \mathbf{S}_\mathbf{AB}\mathbf{C}_\mathbf{A}\mathbf{S}_\mathbf{AB}^{-1} \tag{75c}$$

$$\ldots \qquad \ldots$$

(compare equation (72)), where the similarity transformation matrix $\mathbf{S}_\mathbf{AB}$ relates vectors in $\mathcal{H}_A$ to the corresponding vectors in $\mathcal{H}_B$ (compare equation (73)):

$$\boldsymbol{a}_i^B = \mathbf{S}_\mathbf{AB}\boldsymbol{a}_i^A \tag{76a}$$

$$\boldsymbol{b}_i^B = \mathbf{S}_\mathbf{AB}\boldsymbol{b}_i^A \tag{76b}$$

$$\boldsymbol{c}_i^B = \mathbf{S}_\mathbf{AB}\boldsymbol{c}_i^A \tag{76c}$$

Similarly, the matrix $\mathbf{S_{AC}}$ converts the matrices $\mathbf{A_A}$ to $\mathbf{A_C}$, $\mathbf{B_A}$ to $\mathbf{B_C}$, $\mathbf{C_A}$ to $\mathbf{C_C}$… and the vectors $\boldsymbol{a}_i^A$ to $\boldsymbol{a}_i^C$ , $\boldsymbol{b}_i^A$ to $\boldsymbol{b}_i^C$ … by equations analogous to equations (75) and (76) respectively.

At first sight it might appear too much of a generalization to allow a similarity transformation from one Hilbert space to another. But the matrices that map vectors *between two Hilbert spaces* are in fact identical to perfectly respectable matrices that act purely *within one Hilbert space*. The key to this realization is the important "parallel axes" relationship

$$\boldsymbol{a}_i^A = \boldsymbol{b}_i^B = \boldsymbol{c}_i^C = \cdots , \; i = 1, 2, 3 \ldots \tag{77}$$

noted in equation (53), which expresses nothing more than the fact that the respective coordinate axes of Hilbert spaces $\mathcal{H}_A, \mathcal{H}_B, \mathcal{H}_C$… are the eigenvectors of $\mathbf{A_A}, \mathbf{B_B}, \mathbf{C_C}$ ... respectively and are *chosen to be parallel*. Starting with the definition of $\mathbf{S_{AB}}$ in equation (76a) and inserting $\boldsymbol{a}_i^A = \boldsymbol{b}_i^B$ to give

$$\boldsymbol{a}_i^B = \mathbf{S_{AB}}\boldsymbol{a}_i^A = \mathbf{S_{AB}}\boldsymbol{b}_i^B = \mathbf{M_{ba}}\boldsymbol{b}_i^B \tag{78}$$

we see that the matrix $\mathbf{S_{AB}}$, which acts *between* Hilbert spaces to carry the vector $\boldsymbol{a}_i^A$ in $\mathcal{H}_A$ into the corresponding vector $\boldsymbol{a}_i^B$ in $\mathcal{H}_B$, is *the same* as the matrix $\mathbf{M_{ba}}$, which acts purely *within* a single Hilbert space to map a vector $\boldsymbol{b}_i^B$ into a vector $\boldsymbol{a}_i^B$

$$\mathbf{S_{AB}} = \mathbf{M_{ba}} \tag{79}$$

This matrix, or rather its inverse, $\mathbf{M_{ab}} = \mathbf{M_{ba}^{-1}}$, is already familiar from equation (31) in our treatment of the Arrow system, and the form of $\mathbf{S_{BA}} = \mathbf{S_{AB}^{-1}} = \mathbf{M_{ab}}$ is simply that of $\mathbf{M_{ab}}$ in equations (32) and (36).

We now remind ourselves of the reason for using two Hilbert spaces, namely that when the variables A and B are incompatible, the vectors $\boldsymbol{b}_i^A$ are in general non-orthogonal in $\mathcal{H}_A$ because they are related to the vectors $\boldsymbol{a}_i^A$ by a *non-rigid* rotation, effected by a matrix $\mathbf{M_{ab}}$,

$$\boldsymbol{b}_i^A = \mathbf{M_{ab}}\boldsymbol{a}_i^A \tag{80}$$

in which $\boldsymbol{a}_1^A$ and $\boldsymbol{a}_2^A$ are rotated by *different* angles to give $\boldsymbol{b}_1^A$ and $\boldsymbol{b}_2^A$ respectively, as in equations (32) and (33) for the Arrow; this non-orthogonality problem does not arise if the variables A and B are *compatible*, because the value state $(a_i)$ of A and the value state $(b_j)$ of B are then the same state $(a_i, b_j)$, so there is *no* rotation involved. The reason for the non-rigidity of the rotation (80) is that the form of the probabilities $\mathrm{P}_{(b_i)}(a_j) = \left|\left\langle a_j^A \middle| b_i^A \right\rangle\right|^2$ in GIV theories can be wide-ranging, restricted only by the basic rules of probability (namely exclusivity, normalization, etc. – see Section 2) for general incompatible variables A and B. But in QM the incompatible variables A and B are symmetries of space-time, which imposes more severe restrictions on the probabilities. The result (as we show next) is that the rotation in equation (80) becomes a *rigid* rotation, in which $\boldsymbol{a}_1^A$ and $\boldsymbol{a}_2^A$ are rotated by *the same* angles to give $\boldsymbol{b}_1^A$ and $\boldsymbol{b}_2^A$ , so that the vectors $\boldsymbol{b}_i^A$ are now orthogonal in $\mathcal{H}_A$ as well as in $\mathcal{H}_B$. Furthermore, since $\mathbf{M_{ab}} = \mathbf{S_{BA}}$, this means that $\mathbf{S_{BA}}$ will then also effect a *rigid* rotation in which the vectors $\boldsymbol{b}_1^B$ and $\boldsymbol{b}_2^B$ , and all other pure state vectors $\boldsymbol{\psi}^B$ in $\mathcal{H}_B$, are rotated by the *same* angle to give the corresponding vectors $\boldsymbol{b}_1^A$ , $\boldsymbol{b}_2^A$ and $\boldsymbol{\psi}^A$ in $\mathcal{H}_A$, so that the two Hilbert spaces $\mathcal{H}_A$ and $\mathcal{H}_B$ become superimposable (and therefore equivalent) for all pure states, making separate Hilbert spaces unnecessary.

The key to recovering the single Hilbert space formalism of standard QM from the multi-Hilbert space formalism of GIV theory is therefore the requirement that the matrix $\mathbf{S_{BA}} = \mathbf{S_{AB}^{-1}}$ be *unitary*, in order to effect a *rigid* rotation between all pure state vectors in $\mathcal{H}_B$ and the corresponding vectors in $\mathcal{H}_A$. But in the general case, $\mathbf{S_{BA}}$ is *not* unitary. So what, then, does it take to make $\mathbf{S_{BA}}$ unitary, so that a single Hilbert space can be used? We have already seen that *symmetry operators* are unitary, but the matrix $\mathbf{S}$ that effects a similarity transformation between two equivalent irreps as in equations (75) is not necessarily itself a symmetry operation of the group. There are some cases where the $\mathbf{S}$ matrices *are* symmetry operators, as in the case of symmetry operations in the same class, such as rotations about the *x, y* and *z* axes in the full rotation group $R_3$, which are related in a similarity transformation by the symmetry operation $C_3$. But some of the most interesting incompatible variables are definitely *not* related by symmetry, such as position (the approximate generator of Lorentz boosts) and momentum (the generator of spatial translations). For variables such as these, the $\mathbf{S}$ matrices are not symmetry operators, so we cannot use that to infer their unitary.

But although position and momentum are not *related* by symmetry, they are *themselves* symmetries (actually the generators thereof). And when incompatible variables A and B are themselves symmetries (or generators thereof), we can always use the fact that *one* of the similarity transformation equations (75) is a *diagonalization* of a *unitary* matrix, and the matrix that diagonalizes a unitary operator is always itself unitary [22]. Specifically equation (75b), which we re-write here as

$$\mathbf{B_B} = \mathbf{S_{AB}}\mathbf{B_A}\mathbf{S_{AB}^{-1}} \tag{81}$$

is a diagonalization of the unitary matrix $\mathbf{B_A}$ (unitary because it is a symmetry operator) to give the diagonal unitary matrix $\mathbf{B_B}$, so the matrix $\mathbf{S_{AB}}$ must be unitary even if it is not itself a symmetry operator. Similarly, all the other $\mathbf{S}$ matrices $\mathbf{S_{AC}}, \mathbf{S_{BC}}$ … are also unitary. This diagonalization argument – which can be used only if the *variables* are symmetries (or the Hermitian generators of symmetries) – *guarantees* that the $\mathbf{S}$ matrices, and thence (from equation (79)) the corresponding $\mathbf{M}$ matrices, are unitary (even in cases where $\mathbf{S}$ is not a symmetry operator) thus making the value states of both A and B orthogonal in both $\mathcal{H}_A$ and $\mathcal{H}_B$, and enabling the pure states of the two Hilbert spaces with only restricted Born Rules to be combined into one Hilbert space with an unrestricted Born Rule, as in standard QM. Mixed states in the GIV theory can be treated in the same way as in standard QM, i.e. as a classical mixture of pure states. The details are slightly more complicated than for pure states, but the result is the same, namely that a single Hilbert space can be used when the two incompatible variables A and B are symmetries.

Equations such as (81) are valid whether or not the variables A and B are symmetries: if the variables are *not* symmetries, equation (81) is simply a transformation relating matrix operators in different Hilbert spaces, but if the variables *are* symmetries, it becomes, for any elementary system, a generalized equivalence relation between matrix representatives in irreps $\Gamma_A$ and $\Gamma_B$ of the symmetry group of the system in Hilbert spaces $\mathcal{H}_A$ and $\mathcal{H}_B$ respectively. So why can't the diagonalization argument be used when the variables are *not* symmetries? Because unless the variables are symmetries, there is in general no reason to suppose that any of the matrices being diagonalized are unitary (although there may be cases of "accidental" unitarity), so there is in general no guarantee that any of the **S** matrices that do the diagonalization are unitary.

In fact, one cannot in general expect the matrix **S** to be unitary unless at least *some* symmetry is present, either in the form of the incompatible variables being *related* by symmetry (in which case **S** is a symmetry operator and therefore unitary), or in the form of the incompatible variables *themselves* being symmetries (in which case **S** is unitary because it diagonalizes a unitary operator). And we have seen that **S** *must* be unitary in order for the incompatible variables to be representable in the same Hilbert space so that the familiar *unrestricted* Born Rule can be used. So the Born Rule relies on at least *some* symmetry being present. The classical Arrow system is a case in which the incompatible variables A and B are not themselves symmetries but are *related* by rotational symmetry (isotropy of space), i.e. the corresponding operations are in the *same class*, thus allowing them to be represented in the same Hilbert space. In QM *some* of the incompatible variables are in the same class, but for the others that are *not* in the same class it is only the fact that they are *themselves* symmetries that makes them representable in the same Hilbert space thus enabling the unrestricted Born Rule to be used. Specifically, components $\hat{J}_q$ of angular momentum about different axes are *related* by the rotational symmetry of space and are therefore in the same class (as well as being *themselves* symmetries as generators of rotational symmetry); but position and momentum are *not related* by symmetry and so are *not* in the same class, and it is only the fact that they are *themselves* symmetries that enables them to be represented in the same Hilbert space, thus allowing the familiar Born Rule to be used.

## 6 The "Born postulate" from symmetry

We have thus shown that QM is a special case of GIV theories in which *all* the fundamental variables are *themselves* symmetries. Some of these symmetries are incompatible, leading to the "commutator postulate", which is no longer a postulate because it follows automatically from the Poincaré symmetry of spacetime, as shown in Section 3. And we have now shown that the "Born postulate" is also no longer a postulate: although it is not *primarily* derived from symmetry – rather it is simply a free Pythagorean *construction* for accommodating basic features of classical probability theory in Hilbert spaces – it *is* the fact that the fundamental variables of QM are *themselves* symmetries that allows them to be represented in a *single* Hilbert space, thus enabling the *restricted* Born Rule to take on its familiar *unrestricted* form, in agreement with Gleason's theorem. So both the main "postulates" of QM are no longer postulates because they follow automatically from the Poincaré symmetry of spacetime, once the incompatible *variables* of QM are recognized as incompatible *symmetries*. Thus QM truly does come from symmetry!

## 7 Uncertainty and interference in classical systems

We now use our GIV theory to show that *any* probabilistic system – "quantal" or classical – that has incompatible variables will obey an *uncertainty principle* and exhibit *interference*. We demonstrate this for the general case of two incompatible variables, each of which has just two possible measurement outcomes, $a_1$ and $a_2$ for A, and $b_1$ and $b_2$ for B, as in equations (46) to (51), recalling that in GIV theory one must use Hilbert space $\mathcal{H}_A$ to calculate the probability of A outcomes, and Hilbert space $\mathcal{H}_B$ to calculate the probability of B outcomes.

We take *uncertainty principle* to have the general meaning that *certainty about the anticipated result of an experiment to measure a variable A can only be bought at the expense of uncertainty in the anticipated result of an experiment to measure another variable B with which it is incompatible* [5]. For incompatible variables A and B, the result of an experiment to measure A when in state $(a_1)$ is *certain*, always giving the outcome $a_1$ and never the outcome $a_2$.

$$\mathrm{P}_{(a_1)}(a_1) = |\langle a_1^A | a_1^A \rangle|^2 = 1 \tag{82a}$$

$$\mathrm{P}_{(a_1)}(a_2) = |\langle a_2^A | a_1^A \rangle|^2 = 0 \tag{82b}$$

but the result of an experiment to measure B when in state $(a_1)$ is *uncertain*, yielding either $b_1$ or $b_2$, not with certainty but with *probabilities*

$$\mathrm{P}_{(a_1)}(b_1) = |\langle b_1^B | a_1^B \rangle|^2 \tag{83a}$$

$$\mathrm{P}_{(a_1)}(b_2) = 1 - |\langle b_1^B | a_1^B \rangle|^2 \tag{83b}$$

This uncertainty arises because common eigenstates of the form $(a_i, b_j)$ simply do not exist for incompatible variables. By contrast, if A and B are compatible, a complete set of common eigenstates of A and B *does* exist, so we then have certainty for both variables,

$$\mathrm{P}_{(a_1,b_1)}(a_1) = 1,\ \mathrm{P}_{(a_1,b_1)}(a_2) = 0 \qquad (84a)$$

$$\mathrm{P}_{(a_1,b_1)}(b_1) = 1,\ \mathrm{P}_{(a_1,b_1)}(b_2) = 0 \qquad (84b)$$

Uncertainty can arise even in purely classical systems such as our Arrow, for which states such as $(a_+, b_+)$ do not exist because the Arrow cannot be pointing in both the A and B directions at the same time.

The term *interference* has two related meanings. We focus first on the meaning in the context of a pair of incompatible variables A and B, where it means that measurement of B can *interfere with* (affect the outcome of) an experiment to measure A. We consider first, for purposes of comparison, what happens in the case of compatible variables. If we perform a *direct* measurement of the variable A on the pure state $\psi = (a_1, b_1)$, we obtain the anticipated result A = $a_1$ with certainty,

$$(a_1, b_1) \xrightarrow{\text{Observe A}} (a_1, b_1),\ \text{Outcome A} = a_1 \qquad (85a)$$

If we perform an *indirect* measurement of A on state $(a_1, b_1)$, in which we first observe B and then observe A, we also obtain the expected outcome A = $a_1$ with certainty,

$$(a_1, b_1) \xrightarrow{\text{Observe B}} (a_1, b_1) \xrightarrow{\text{Observe A}} (a_1, b_1),\ \text{Outcome A} = a_1 \qquad (85b)$$

We see that measuring B before measuring A does not interfere with the measurement of A: the same outcome, $\mathrm{A} = a_1$, is obtained in both direct and indirect measurements, reflecting that fact that the operators for compatible variables commute, $[\hat{A}, \hat{B}] = 0$. Turning now to the case of incompatible variables, and considering measurements of the variable A on an arbitrary state ψ, either directly, or via a prior measurement of B, we obtain the following diagrammatic equations,

$$\psi \xrightarrow{\text{Observe A}} \begin{cases} (a_1) \\ (a_2) \end{cases} \qquad (86a)$$

$$\psi \xrightarrow{\text{Observe B}} \begin{cases} (b_1) \xrightarrow{\text{Observe A}} \begin{cases} (a_1) \\ (a_2) \end{cases} \\ \\ (b_2) \xrightarrow{\text{Observe A}} \begin{cases} (a_1) \\ (a_2) \end{cases} \end{cases} \qquad (86b)$$

noting that observation of B causes a transition from the state (ψ) to an eigenstate $(b_i)$ of B, and observation of A then causes a transition to an eigenstate $(a_i)$ of A. For the *direct* measurement in (86a) the probability of obtaining the result $a_i$ from a starting state ψ is given by

$$\mathrm{P}_\psi^{\mathrm{dir}}(a_i) = \left|\langle a_i^A | \psi^A \rangle\right|^2, \;\; i = 1,2 \qquad (87a)$$

The *indirect* measurement of A in (86b) yields the result $a_i$ with a probability $\mathrm{P}_\psi^{\mathrm{indir}}(a_i)$ given by the sum of the two sub-processes via $b_1$ and $b_2$ respectively:

$$\mathrm{P}_\psi^{\mathrm{indir}}(a_i) = \mathrm{P}_{b_1}(a_i)\mathrm{P}_\psi(b_1) + \mathrm{P}_{b_2}(a_i)\mathrm{P}_\psi(b_2)$$
$$= \left|\langle a_i^A | b_1^A \rangle\right|^2 |\langle b_1^B | \psi^B \rangle|^2 + \left|\langle a_i^A | b_2^A \rangle\right|^2 |\langle b_2^B | \psi^B \rangle|^2 \qquad (87b)$$

The probability $\mathrm{P}_\psi^{\mathrm{dir}}(a_i)$ of obtaining $a_i$ by the direct measurement certainly has a different algebraic form from the probability $\mathrm{P}_\psi^{\mathrm{indir}}(a_i)$ of obtaining $a_i$ by the indirect measurement, but it is not in general immediately obvious that their numerical values differ until one inserts values for the individual amplitudes for a specific case. A convenient specific case is provided by considering the probability of obtaining the outcome $\mathrm{A} = a_2$ from a starting state $(\psi) = (a_1)$, which must be zero in a direct measurement,

$$\mathrm{P}_{(a_1)}^{\mathrm{dir}}(a_2) = |\langle a_2^A | a_1^A \rangle|^2 = 0 \qquad (88a)$$

but cannot also be zero in an indirect measurement,

$$\mathrm{P}_{(a_1)}^{\mathrm{indir}}(a_2) = |\langle a_2^A | b_1^A \rangle|^2 \, |\langle b_1^B | a_1^B \rangle|^2 + |\langle a_2^A | b_2^A \rangle|^2 \, |\langle b_2^B | a_1^B \rangle|^2 \neq 0 \qquad (88b)$$

because this would require at least some of the matrix elements in (88b) to be zero, and it is readily shown that this is not possible if the variables A and B are incompatible. For example, $\langle a_2^A | b_1^A \rangle = 0$ would require $|b_1^A\rangle = |a_1^A\rangle$ (to within a phase factor), which can only be the case if A and B are compatible, and similar considerations apply to all the other matrix elements.

The inequality of $\mathrm{P}_\psi^{\mathrm{dir}}(a_i)$ and $\mathrm{P}_\psi^{\mathrm{indir}}(a_i)$ becomes more obvious in cases where *symmetry* is present. Applying the general equations (87) to our classical Arrow system, for which the index $i$ is + or $-$, and considering just the probability of obtaining the result $a_+$, we have

$$\mathrm{P}_\psi^{\mathrm{dir}}(a_+) = |\langle a_+^A | \psi^A \rangle|^2 \qquad (89)$$

for the direct measurement. We now make use of the fact that in the Arrow the eigenstates $(a_+)$ and $(a_-)$ of A are related to one another by $C_2$ symmetry, as are the eigenstates $(b_+)$ and $(b_-)$ of B: we saw in equations (28) that this symmetry results in the eigenstates of B being orthogonal not only in Hilbert space $\mathcal{H}_B$ but also in Hilbert space $\mathcal{H}_A$, so they form a complete orthonormal set in $\mathcal{H}_A$, allowing us to insert the closure relation

$$|b_+^A\rangle\langle b_+^A| + |b_-^A\rangle\langle b_-^A| = 1 \qquad (90)$$

into $\langle a_+^A | \psi^A \rangle$ in equation (89), giving

$$\mathrm{P}_\psi^{\mathrm{dir}}(a_+) = |\langle a_+^A | b_+^A \rangle\langle b_+^A | \psi^A \rangle + \langle a_+^A | b_-^A \rangle\langle b_-^A | \psi^A \rangle|^2 \qquad (91)$$

Since the $C_2$ symmetry of the Arrow makes the eigenstates of B orthogonal in $\mathcal{H}_A$ as well as in $\mathcal{H}_B$, a single Hilbert space can now be used with the additional assumption that space itself is isotropic (see the discussion following equations (40)), so we drop the superscripts to give

$$\mathrm{P}_{\psi}^{\mathrm{dir}}(a_+) = |\langle a_+|b_+\rangle\langle b_+|\psi\rangle + \langle a_+|b_-\rangle\langle b_-|\psi\rangle|^2 \tag{92a}$$

$$\mathrm{P}_{\psi}^{\mathrm{indir}}(a_+) = |\langle a_+|b_+\rangle|^2\,|\langle b_+|\psi\rangle|^2 + |\langle a_+|b_-\rangle|^2\,|\langle b_-|\psi\rangle|^2 \tag{92b}$$

for the direct and indirect measurements in the Arrow. It is now much clearer that $\mathrm{P}_{\psi}^{\mathrm{dir}}(a_+) \neq \mathrm{P}_{\psi}^{\mathrm{indir}}(a_+)$, since equations (92) are of the form

$$\mathrm{P}_{\psi}^{\mathrm{dir}}(a_+) = |R_1 + R_2|^2 = R_1^2 + R_2^2 + 2R_1R_2 \tag{93a}$$

$$\mathrm{P}_{\psi}^{\mathrm{indir}}(a_+) = R_1^2 + R_2^2 \tag{93b}$$

which is reminiscent of the second meaning of interference, in the context of waves overlapping in phase or out of phase to produce constructive or destructive interference. This type of interference is said to occur in wave optics when *the interaction of two light beams yields a resultant irradiance that deviates from the sum of the component irradiances* [23]. The irradiance at a given location is proportional to the probability of detecting a photon at that location, which is proportional to the *square* of the *amplitude* of the electric field of the light detected at that location. Interference occurs simply because *amplitudes are additive*, resulting in the *square of the sum* being different from the *sum of the squares*. The use of the term "amplitude" for the *square root* of a probability may give the impression that interference only occurs for systems that are literally physical waves, since square roots can be positive or negative like the amplitude of the peaks and troughs of a wave. But interference can actually occur in *any* probabilistic system, wave-like or not, that has incompatible variables, and we follow Kirkpatrick [1] in broadening the wave optics definition of interference in terms of resultant irradiance deviating from the sum of component irradiances to the more general statement that interference occurs when *the probability of a process deviates from the sum of the probabilities of the sub-processes*, which is exactly what we see in the inequality of the results of direct and indirect measurements in our classical Arrow system in equations (92) and (93).

Richard Feynman famously said [24] that QM is about "adding amplitudes", to which one might add "*before* squaring them", because it is "adding amplitudes *before* squaring them" that is responsible for the wave-like interference effect that has hitherto been considered the hallmark of QM. But we see from equations (93) that there is in fact a direct parallel between interference in the probability patterns of our purely classical Arrow system and corresponding probability patterns in the famous double-slit electron diffraction experiments that gave birth to the QM notion of wave-particle duality. It is therefore clear that although interference ("adding amplitudes") is a prominent feature of QM, it is not unique to QM and is not necessarily indicative of actual physical waves, rather it is a general feature of all probability theories with incompatible variables. However, interference is particularly easy to *recognize* in QM, because what distinguishes QM from other probability theories is the fact that *symmetry* enables all the fundamental variables to be represented by *orthogonal* axes in a single Hilbert space, which in turn allows use of a closure relation to arrive at expressions of the form (93a), in which *probabilities* show interference even in the absence of actual physical waves. Symmetries in the double-slit system, for example, lead to an enhanced *impression* of wave character, as we will discuss in more detail in a later paper.

## 8 Conclusion

We have thus shown that, surprising though it may seem, incompatible variables – with concomitant uncertainty and interference – are indeed not the exclusive preserve of QM, nor are they on a fundamental level the consequences of wave character or quantization. We add to the growing list of classical systems with incompatible variables [1, 2] our "Arrow" system, a multi-variable analogue of the coin toss that shows uncertainty and interference similar to that in double-slit diffraction experiments. Inspired by Weinberg's suggestion [3] that it would be useful to find a larger, more general theory in which QM appears as a special case, we have constructed a *general incompatible variables* (GIV) theory, to describe *any* system with incompatible variables, and we find that what distinguishes QM from other GIV theories is *symmetry*: the incompatible variables of QM are simply *incompatible symmetries*. Specifically, the fundamental variables of QM – position, linear momentum, energy and angular momentum – are replaced at a deeper level by, respectively, Lorentz boost symmetries, spatial translation symmetries, time translational symmetry, and rotational symmetries of the Poincaré group.

Considering the two central postulates of QM, the role of Poincaré symmetry in the "commutator postulate" $[\hat{x}, \hat{p}_x] = i\hbar$, although not as widely appreciated as it should be, is abundantly clear [7,14], as discussed in Section 3, but the origin of the "Born postulate" $P_\psi(o_i) = |\langle o_i|\psi\rangle|^2$ has hitherto remained obscure, despite Gleason's theorem [8] showing that the Born Rule is the only probability rule that works if all variables are represented in a single Hilbert space. However, Hughes [5] realized that two *incompatible* variables A and B *cannot in general* be represented in just one Hilbert space, because the vectors representing the eigenstates of B, while orthogonal in a Hilbert space $\mathcal{H}_B$ (in which the eigenvectors of B form the axes) are *not in general* (except in very special cases) orthogonal in a Hilbert space $\mathcal{H}_A$ (in which the eigenvectors of A form the axes), and, so far as we are aware, Hughes is the only one to have noticed this.

But the central point of this paper, which has not, to our knowledge, been noticed by anyone until now, is that *symmetries* allow incompatible variables to be represented together in a single Hilbert space, as we highlight using our simple classical Arrow system.

We follow Hughes [5] in using a separate Hilbert space for calculating the outcomes of a measurement of each incompatible variable in our GIV theory, and we introduce the new concept of a *restricted* Born Rule $\mathrm{P}_{\psi}(o_i) = |\langle o_i|\psi\rangle|^2$ in which $|o_i\rangle$ is restricted to being one of the orthonormal *axes* (i.e. a basis vector) of that particular Hilbert space, in contrast to the usual *unrestricted* Born Rule, in which not only $|\psi\rangle$ but also $|o_i\rangle$ can be *any* vector within the Hilbert space. We then show that if the fundamental variables are *symmetries*, as in QM, the many Hilbert spaces of GIV theory are then related to one another by *rigid* transformations (as opposed to non-rigid transformations in the general case), which allows the many Hilbert spaces to be superimposed and combined into one single Hilbert space, with a single, unrestricted Born Rule for all of the variables.

QM is thus a special case of GIV theories, one in which all the fundamental variables are symmetries. The two central postulates of QM are no longer "underivable", but in fact come from these symmetries, some of which are incompatible. The derivation of the "commutator postulate" from Poincaré symmetry has been presented in the most detail by Bohr & Ulfbeck [7]. But there has hitherto been no truly satisfactory "derivation" of the "Born postulate", and the attempt by Bohr & Ulfbeck [7] to derive it too from symmetry did not succeed. However, our work in this paper now provides a much clearer understanding of the Born Rule: although it is not *primarily* derived from symmetry – rather it is simply a free Pythagorean *construction* for accommodating basic features of classical probability theory in Hilbert spaces – it *is* the Poincaré symmetry of the states that allows the fundamental variables of QM to be represented in a *single* Hilbert space, thus enabling the *restricted* Born Rule to take on its familiar *unrestricted* form, in agreement with Gleason's theorem.

In summary, we find that when the variables are taken to be the symmetries of the Poincaré group (or their generators), the two central postulates of QM – the "commutator postulate" and the "Born postulate" – follow automatically, and are therefore no longer postulates. QM thus emerges as a special case of GIV theories for which the variables are symmetries, specifically (for the case of non-interacting particles or fields) symmetries of the Poincaré group of spacetime. So QM does truly come from symmetry!

In later papers we will provide a diagrammatic understanding of the commutator $[\hat{x}, \hat{p}_x]$, analogous to the more familiar understanding of the commutator $[\hat{J}_x, \hat{J}_y]$, discuss further the enhanced impression of wave character in the double-slit experiment, and consider the implications of our conclusions for a new interpretation of QM as a theory of incompatible symmetries. Future work could extend our discussion of GIV systems to include fields, interactions, composite systems, and entanglement.

## Appendix 1 The Born "postulate" as a free Pythagorean construction

In this Appendix we examine the nature of the Born "postulate" or Born Rule

$$\mathrm{P}_{\psi}(o_i) = |\langle o_i|\psi\rangle|^2 = \cos^2 \omega_{\psi o_i} \tag{A1}$$

where $\omega_{\psi o_i}$ is the Pythagorean angle between $|\psi\rangle$ and $|o_i\rangle$ in Hilbert space, and we conclude that the Born Rule is not really a "postulate" as such, neither is it really quantum mechanical in content, but rather is best seen as a free Pythagorean *construction* for accommodating basic features of *classical* probability theory in Hilbert spaces. One can always *choose* to use the Born Rule $\mathrm{P}_{\psi}(o_i) = |\langle o_i|\psi\rangle|^2$ as a construction (since the construction relies only on the truth of Pythagoras' theorem), *provided* that $|\psi\rangle$ and $|o_i\rangle$ are vectors in the same Hilbert space and that the eigenvectors $|o_i\rangle$ form a complete orthonormal set in that space.

The Born Rule was born when Born proposed his $|\psi(x)|^2$ probability density for finding a particle at a position *x* by simple physical analogy with Einstein's interpretation of the square of the amplitude of an electromagnetic wave as the probability of finding a photon. It may have been little more than an inspired guess on Born's part, especially in view of the fact that his original 1926 paper [25] proposed $\psi(x)$ as the probability density, which he changed to $|\psi(x)|^2$ only as a footnote in the proof stage. But it certainly seems to have been a very good guess, as the validity of the $|\psi(x)|^2$ Born Rule has since been completely borne out by experiment, including by recent high precision interference tests [26]. Furthermore, the Born Rule was put on a more secure mathematical foundation in 1957 by Gleason's theorem [8], which shows that the Born expression is the *only* possible rule for calculating probabilities in a theory that represents outcomes of measurements as orthonormal vectors in a single Hilbert space. But not all were satisfied with Gleason's purely mathematical derivation, and many attempts were made to find more physical proofs based on deep QM principles, usually in the context of particular interpretations of QM, most notably by Deutsch [10] and Wallace [11], based on the many-worlds interpretation, and by Zurek [12], based on decoherence by entanglement with the environment.

However, the fact that Gleason was able to arrive at the Born Rule as a purely mathematical result pertaining to vectors and operators in Hilbert space does raise suspicion that the Born Rule does not come from any deeper QM principles, but simply arises naturally in Hilbert space formulations of *classical* probability theories. Hilbert spaces (using the orthonormal eigenstates of Hermitian operators as basis vectors) tend to be associated with QM, but it is less widely known that a Hilbert space formulation of classical mechanics was developed in 1931 by Koopman [27]. Indeed, the whole operator-eigenstate-eigenvalue formalism is not unique to QM either, and Brumer and Gong [13] use this formalism within Hilbert space to derive an analogue of the Born Rule in classical mechanics that deserves to be more widely appreciated, confirming as it does that the Born Rule – wherever it may come from – is definitely not unique to QM.

Hilbert spaces, with their orthogonal axes, are intrinsically Pythagorean in nature, thus allowing for the *normalization* and *exclusivity* features of probability theories. In our treatment of the Arrow and GIV theories in Sections 4 and 5 we need to construct Hilbert space vectors from probabilities, so we briefly review here the familiar Pythagorean construction used to do this in QM.

Consider a general system with just two eigenstates, $|1\rangle$ and $|2\rangle$, in which the probabilities $\mathrm{P}_\psi(1)$ and $\mathrm{P}_\psi(2)$ of obtaining respectively result 1 and result 2 when making a measurement on the system in some starting state $|\psi\rangle$ are given by

$$\mathrm{P}_\psi(1) = f \qquad \text{(A2a)}$$

$$\mathrm{P}_\psi(2) = 1 - f \qquad \text{(A2b)}$$

with equation (A2b) following from (A2a) if the probabilities are normalized. One can always *choose* to represent *any* real function $f$ with $|f| \leq 1$ (probability or otherwise) as $\cos^2 \omega$, in which case the Pythagorean identity $\cos^2 \omega + \sin^2 \omega = 1$ and the trigonometric identity $\sin \omega = \cos\left(\frac{\pi}{2} - \omega\right)$ enable equations (A2) to be written as

$$\mathrm{P}_\psi(1) = \cos^2 \omega \qquad \text{(A3a)}$$

$$\mathrm{P}_\psi(2) = \sin^2 \omega = \cos^2\left(\frac{\pi}{2} - \omega\right) \qquad \text{(A3b)}$$

where the "Pythagorean angle" $\omega$ is given by

$$\omega = \cos^{-1} f^{1/2} = \tan^{-1} \frac{(1-f)^{1/2}}{f^{1/2}} \qquad \text{(A4)}$$

Since $\cos \omega$ can be identified with the scalar product of two normalized vectors that are related by an angle $\omega$ in a Hilbert space, equations (A3) can then be written as

$$\mathrm{P}_\psi(1) = \cos^2 \omega_{\psi 1} = |\langle 1|\psi\rangle|^2 \qquad \text{(A5a)}$$

$$\mathrm{P}_\psi(2) = \cos^2\left(\frac{\pi}{2} - \omega_{\psi 1}\right) = \cos^2 \omega_{\psi 2} = |\langle 2|\psi\rangle|^2 \qquad \text{(A5b)}$$

where $\omega_{\psi 1}$ is the angle relating $|\psi\rangle$ to $|1\rangle$, and $\omega_{\psi 2}$ is the angle relating $|\psi\rangle$ to $|2\rangle$. We see that expressing probabilities in terms of the Pythagorean angle $\omega$ to ensure *normalization* of the probabilities (as in equations (A3)) also demands orthogonality of the states $|1\rangle$ and $|2\rangle$, i.e. it demands that $\omega_{\psi 2} = \frac{\pi}{2} - \omega_{\psi 1}$, as in equation (A5b) (since Pythagoras' theorem applies only to right-angled triangles), thus also ensuring *exclusivity*, i.e. a system in state $|1\rangle$ cannot also be in state $|2\rangle$.

Expressing probabilities in "cosine-squared" form as in equations (A5) is thus not a unique feature of QM, and it is not a *postulate* as such, rather it is a *Pythagorean construction* to accommodate basic features of *classical* probability theory, namely *normalization* and *exclusivity*; and the Born-type "scalar product squared" form is just the Pythagorean construction expressed in a form suitable for representing classical or QM probabilities in Hilbert space. In fact, the very essence of a Hilbert space spanned by an orthonormal basis is its Pythagorean nature [5], with its normalized vectors and orthogonal axes.

In a two-dimensional Hilbert space spanned by the basis $\{|1\rangle, |2\rangle\}$, the normalized state $|\psi\rangle$ can be written (to within arbitrary phase factors) as

$$|\psi\rangle = f^{1/2}|1\rangle + (1-f)^{1/2}|2\rangle \quad \text{(A6)}$$

or

$$|\psi\rangle = \cos\omega\,|1\rangle + \sin\omega\,|2\rangle \quad \text{(A7)}$$

where $\omega$ is the Pythagorean angle relating $|\psi\rangle$ and $|1\rangle$ in Hilbert space, given by (A4). In Sections 4 and 5 we use equation (A6) to construct the vectors $|\psi\rangle$ in Hilbert space from the probabilities $f$. We illustrate this here with the classical example of the familiar coin toss experiment, in which $|1\rangle$ and $|2\rangle$ correspond respectively to heads, $|H\rangle$, and tails, $|T\rangle$. The Hilbert space vector $|\psi\rangle$ representing an "honest" coin, which has equal probability of landing heads or tails, can be represented in Hilbert space by a vector $|\psi\rangle$ of unit length oriented at 45° to both axes. The projection $\langle H|\psi\rangle$ of $|\psi\rangle$ onto the H axis is then $1/\sqrt{2}$, as is the projection $\langle T|\psi\rangle$ of $|\psi\rangle$ onto the T axis. The squares of these projections are then seen to be the observed probabilities,

$$P_\psi(H) = |\langle H|\psi\rangle|^2 = \cos^2\frac{\pi}{4} = ½ \quad \text{(A8a)}$$

$$P_\psi(T) = |\langle T|\psi\rangle|^2 = \sin^2\frac{\pi}{4} = ½ \quad \text{(A8b)}$$

and by Pythagoras' theorem the probability rule

$$P_\psi(H) + P_\psi(T) = 1 \quad \text{(A9)}$$

is seen to be satisfied. A weighted or "loaded" coin can be described by a state $|W\rangle$ where the Pythagorean angle $\omega$ made by $|W\rangle$ with the H axis is something other than 45°. We now have

$$P_W(H) = |\langle H|W\rangle|^2 = \cos^2\omega = f \quad \text{(A10a)}$$

$$P_W(T) = |\langle H|T\rangle|^2 = \cos^2\left(\frac{\pi}{2} - \omega\right) = \sin^2\omega = 1 - \cos^2\omega = 1 - f \quad \text{(A10b)}$$

The Pythagorean construction in equation (A6) then gives the state of a loaded coin as

$$|\mathrm{W}\rangle = f^{½}|\mathrm{H}\rangle + (1-f)^{½}|\mathrm{T}\rangle = \cos\omega\,|\mathrm{H}\rangle + \sin\omega\,|\mathrm{T}\rangle \quad \text{(A11)}$$

So if $\omega = 30^0$, say, we have $\cos\omega = \sqrt{3}/2$, $\sin\omega = ½$, giving $\mathrm{P_W(H)} = ¾$ and $\mathrm{P_W(T)} = ¼$ for a loaded coin that lands heads 75% of the time and tails 25% of the time.

Although the Hilbert space representation of the "honest" coin state

$$|\psi\rangle = \frac{1}{\sqrt{2}}(|\mathrm{H}\rangle + |\mathrm{T}\rangle) \quad \text{(A12)}$$

may *look* rather like some expressions for pure states in quantum mechanics, it is actually a mixed state in classical probability theory for the coin toss. Here we define a *pure state* as a state for which we can predict the outcome of a measurement of at least one variable performed on this state with 100% certainty; any other state is a *mixed state.* In the case of the coin toss, we have a single variable with possible values H and T, and the only pure states are thus $|\mathrm{H}\rangle$ and $|\mathrm{T}\rangle$. For the state $|\mathrm{H}\rangle$, a toss and landing of the coin – that is, a measurement of what one might call the "sidedness" variable – will give the outcome "heads" with 100% certainty (such a coin is weighted to always land "heads"); we leave it to the reader to suggest how the state $|\mathrm{H}\rangle$ might be prepared! But the "honest" coin state $|\psi\rangle$ is a mixed state that lands "heads" 50% of the time and "tails" 50% of the time, and is equivalent to a 50:50 classical mixture of coins in the states $|\mathrm{H}\rangle$ and $|\mathrm{T}\rangle$.

We note that because expressing probabilities in terms of the Pythagorean angle $\omega$ in Hilbert space is just a construction, $\omega$ does not necessarily correspond to any actual *physical* angle. In the coin toss there are *no* physical angles, and the Pythagorean angle $\omega$ is simply an angle in Hilbert space related to the *weighting* of the coin. In our classical "Arrow" system, described in Section 4, there *is* a physical angle $\theta$ describing the orientation of the Arrow, but the probability $\mathrm{P} = f$ of having that orientation does not necessarily equal $\cos^2\theta$, and in general could be almost any function of $\theta$. We can *choose* to use the Pythagorean construction to write $f = \cos^2\omega(\theta)$ where $\omega$ is some function $\omega(\theta)$ of $\theta$, but in classical systems this is a purely artificial construction, and the relationship between the physical angle $\theta$ and the Pythagorean angle $\omega$ may be so complicated that writing the probability as a function of $\omega$ instead of $\theta$ may have no practical use. In QM, however, we find (see Appendix 3) that the Pythagorean angle $\omega$ has a very simple relationship to the physical angle $\theta$, specifically $\omega = ½\theta$ for spin ½ systems, so that equation (A1), the Born postulate, becomes

$$\mathrm{P} = \cos^2\omega = \cos^2 ½\theta \quad \text{(A13)}$$

where $\theta$ is the *physical* angle between two spin orientations, and $\omega$ is the angle between the representation of these orientations in Hilbert space. Very simple and unique results such as this are what distinguishes QM from more general probabilistic theories, and they can be obtained from these more general theories by introducing *symmetry* constraints.

The Born Rule $\mathrm{P}_{\psi}(o_i) = |\langle o_i|\psi\rangle|^2 = \cos^2 \omega_{\psi o_i}$ is thus best seen as a free Pythagorean construction to represent the normalization and exclusivity features of classical or QM probability in a Hilbert space spanned by an orthonormal basis. We are always free to *choose* to use this Pythagorean construction, provided that, firstly, $|\psi\rangle$ and $|o_i\rangle$ belong to the same Hilbert space (since scalar products are only defined for vectors in the same Hilbert space) and, secondly, that the possible measurement outcomes $|o_i\rangle$ form a complete orthonormal set in that Hilbert space (to accommodate the normalization and exclusivity features of probability theory). But Gleason's theorem goes further, proving that the Born Rule is not merely a convenient choice of probability rule but is actually the *only* possible choice if *all* vectors describing the system belong to a single Hilbert space so that *every* set $\{|o_i\rangle\}, \{|o_i'\rangle\}, \ldots$ of eigenvectors – corresponding to *all* the observables $O, O', \ldots$ – is a complete orthonormal set in that Hilbert space.

Gleason's own proof of his theorem [8] is lengthy, and impenetrable to most non-mathematicians, but the "elementary" proof by Cooke, Keane and Moran [28] is somewhat less impenetrable if read alongside the commentary by Hughes [5]. Gleason's theorem itself has been stated in many different ways, but the clearest for our purposes is that provided by Hughes [5] (p.147): "For any measure μ on a sub-space L of a Hilbert space $\mathcal{H}$ of dimension at least 3, there exists a positive self-adjoint (i.e. Hermitian) operator T of the trace class (i.e. an operator that has a trace) such that, for all subspaces L of $\mathcal{H}$,

$$\mu(L) = \mathrm{Tr}(\mathrm{T}\mathrm{P}_{\mathrm{L}}) \quad \text{(A14)}$$

where $\mathrm{P}_{\mathrm{L}}$ is a projection operator onto L." In mathematics, a *measure* on a set assigns a number to each subset of that set, interpreted intuitively as a measure of size (i.e. a generalization of length, area, volume, etc.). Gleason's theorem is extremely general, and there are many different types of "measure" in different branches of mathematics. For a general measure there could be several vectors in the subset L; but if the measure is the *probability* $\mu_{\psi}(o_i)$ of obtaining one particular outcome $o_i$ when making a measurement on a state ψ, the subset L is taken to be just one vector $|o_i\rangle$ in the Hilbert space $\mathcal{H}$, and we re-write Gleason's theorem (A14) as

$$\mu_{\psi}(o_i) = \mathrm{Tr}\left(\mathrm{T}_{\psi}\mathrm{P}_{o_i}\right) \quad \text{(A15)}$$

In general probability theories, $|o_i\rangle$ is a pure state but $|\psi\rangle$ could be a mixed state, as we saw above for the classical coin toss. Since $|o_i\rangle$ is a pure state, the projection operator $\mathrm{P}_{o_i}$ has the form

$$\mathrm{P}_{o_i} = |o_i\rangle\langle o_i| \quad \text{(A16)}$$

But what form does $\mathrm{T}_\psi$ take for the state $\psi$? Gleason's theorem does not specify the form of the operator T (other than restricting it to be of the trace class), but the type of measure constrains the form of T. Two constraints arise when the measure is a *probability*. The first constraint is that the measure on the whole Hilbert space $\mathcal{H}$ must be equal to one: if $|\psi\rangle$ is in the same Hilbert space $\mathcal{H}$ as $|o_i\rangle$, $\psi$ is then definitely somewhere in $\mathcal{H}$, so the probability of measuring an outcome in the Hilbert space $\mathcal{H}$ is unity, giving

$$\mu_\psi(\mathcal{H}) = 1 \qquad \text{(A17)}$$

The second constraint is that the possible measurement outcomes $o_i$ must be represented by a complete orthonormal set $\{|o_i\rangle\}$ to accommodate the normalization and exclusivity features of probability theory,

$$\langle o_i|o_j\rangle = \delta_{ij} \qquad \text{(A18)}$$

These two constraints impose restrictions on the form of $\mathrm{T}_\psi$ which in turn restrict the form of the probability to the Born form. Substituting (A15) into (A17) gives

$$\mu_\psi(\mathcal{H}) = \mathrm{Tr}\big(\mathrm{T}_\psi \mathrm{P}_{\mathcal{H}}\big) = \mathrm{Tr}\big(\mathrm{T}_\psi \textstyle\sum_i \mathrm{P}_{o_i}\big) = \mathrm{Tr}\textstyle\sum_i \mathrm{T}_\psi |o_i\rangle\langle o_i| = \textstyle\sum_{i,j}\langle o_j|\mathrm{T}_\psi|o_i\rangle\,\langle o_i|o_j\rangle$$

and using (A18) we obtain

$$\mu_\psi(\mathcal{H}) = \textstyle\sum_i\langle o_i|\mathrm{T}_\psi|o_i\rangle = \mathrm{Tr}\ \mathrm{T}_\psi \qquad \text{(A19)}$$

Equating (A17) and (A19), we see that

$$\mathrm{Tr}\ \mathrm{T}_\psi = 1 \qquad \text{(A20)}$$

and any operator with unit trace can be written as a density operator

$$\mathrm{T}_\psi = |\psi\rangle\langle\psi| \qquad \text{(A21)}$$

(since all density operators have unit trace provided $\psi$ is normalized). Substituting (A21) and (A16) into (A15) then gives

$$\mu_\psi(o_i) = \mathrm{Tr}\big(\mathrm{T}_\psi \mathrm{P}_{o_i}\big) = \mathrm{Tr}(|\psi\rangle\langle\psi|o_i\rangle\langle o_i|) = \textstyle\sum_j\langle o_j|\psi\rangle\langle\psi|o_i\rangle\langle o_i|o_j\rangle$$

and, using $\langle o_i|o_j\rangle = \delta_{ij}$ from (A18), Gleason's theorem now gives

$$\mu_\psi(o_i) = \langle o_i|\psi\rangle\langle\psi|o_i\rangle = |\langle o_i|\psi\rangle|^2 \qquad \text{(A22)}$$

as per the Born Rule $\mathrm{P}_\psi(o_i) = |\langle o_i|\psi\rangle|^2$.

Demonstrations of the derivation of the Born Rule from Gleason's theorem typically take the set of outcomes $\{o_i\}$ to correspond to the set of eigenstates $\{|o_i\rangle\}$ of the observable $O$ that form the *axes* of the Hilbert space $\mathcal{H}$ [8]. But replacing $\{|o_i\rangle\}$ with *any* complete orthonormal set of vectors in the Hilbert space in the above derivation also leads to the Born Rule: Gleason's theorem (A14) actually allows $|\mathrm{L}\rangle = |o_i\rangle$ to be *any* vector in $\mathcal{H}$, since it refers to "*all* subspaces L". So the Born Rule is often written as $\mathrm{P}_u(v) = |\langle v|u\rangle|^2$, and is normally taken to apply to *any* two pure states $|u\rangle$ and $|v\rangle$ (eigenstates of variables U and V respectively) that can be represented in the same Hilbert space, and the outcome states $\{|v\rangle\}$ are not restricted to those that form the axes of the Hilbert space.

Although Gleason's theorem shows that the Born Rule $\mathrm{P}_u(v) = |\langle v|u\rangle|^2$ is the *only* probability measure that works when $|u\rangle$ and $|v\rangle$ can be represented in the same Hilbert space and the vectors $\{|v\rangle\}$ form a complete orthonormal set, the theorem is only true for Hilbert spaces of dimension three and above. But this does *not* mean that the Born expression for probability *cannot* be used for Hilbert spaces that are only of dimension two (of which there are many in QM, e.g. those representing spin ½ particles), it just means that one *cannot rule out* the existence of other alternative probability measures for two-dimensional spaces (and indeed alternative probability measures *have* been found in two dimensions, most notably by Kadison). This is of no serious consequence, however, because we can demand that QM should use a uniform probability measure that is applicable to variables with *any* number of eigenvalues, and since the Born expression is the only probability measure that works for more than two eigenvalues, it is the only one that should be used.

Bohr and Ulfbeck [7] sought to prove that the Born Rule $\mathrm{P}_u(v) = |\langle v|u\rangle|^2$ is a consequence of the variables U and V being *symmetries*, but they tacitly assumed from the beginning that $|u\rangle$ and $|v\rangle$ are represented in a *single* Hilbert space. This makes their somewhat impenetrable proof a red herring, because Gleason's theorem shows that if $|u\rangle$ and $|v\rangle$ can be represented in a single Hilbert space, the probability $\mathrm{P}_u(v)$ *must* be $|\langle v|u\rangle|^2$, *symmetry or no symmetry*. So once a single Hilbert space is assumed, the fact that the variables are symmetries is *not needed* to obtain the Born expression, because it follows automatically from Gleason's theorem. But the Born Rule *cannot* be used when $|u\rangle$ and $|v\rangle$ are in different Hilbert spaces, as scalar products $\langle v|u\rangle$ are only defined for vectors $|u\rangle$ and $|v\rangle$ that belong to the *same* Hilbert space, and it *cannot* be used if the outcome vectors $\{|v\rangle\}$ are not orthonormal – and therein lies the problem if the variables U and V are *incompatible*. We show in Section 5 that this problem is solved when we introduce the Poincaré symmetry of spacetime, which enables the eigenstates of not only compatible variables but also *incompatible* variables to be represented in the *same* Hilbert space. This in turn enables use of the *unrestricted* Born expression $\mathrm{P}_\psi(o_i) = |\langle o_i|\psi\rangle|^2$ (in which the outcome states $|o_i\rangle$ are *not* restricted to being one of the axes of the Hilbert space) to calculate the probabilities of outcomes $a_i$ or $b_i$ or $c_i$… of measurements of *any* of the incompatible variables A, B, C…. in a single Hilbert space, instead of having to calculate the probability of outcomes of measurement of the variable V only in a separate Hilbert space $\mathcal{H}_V$, using the *restricted* Born expression $\mathrm{P}_\psi^{\mathrm{V}}(v_i^V) = |\langle v_i^V|\psi^{\mathrm{V}}\rangle|^2$(in which the outcome states are restricted to being one of the axes of the Hilbert space). This result is consistent with Gleason's theorem, according to which the unrestricted Born expression $P_\psi(o_i) = |\langle o_i|\psi\rangle|^2$ is the *only* possible probability expression that can be used when the theory can be formulated in a *single* Hilbert space.

## Appendix 2 The Haag-Wigner condition for symmetry transformations

Symmetry transformations in physical space (such as translation in a homogeneous space or rotation in an isotropic space) by definition preserve *lengths*, and also *angles* (meaning they are *rigid* transformations), and they also preserve the results of experiments, including the *probabilities* of all outcomes, thus leaving the system *looking* "as if nothing happened". This becomes clearer if instead of *actively* translating or rotating the system we do nothing at all to the system and instead consider the equivalent procedure of changing the coordinates. In this *passive* view it is actually true that "nothing happened" to the system, and the symmetry transformation is simply a "change of view" [2], i.e. a change of observer. We then see that symmetry transformations are *covariant*, meaning that the equations for the transformations have the same form for observers in different reference frames.

The quantities preserved in symmetry transformations − lengths, angles, and probabilities − are all essentially *scalar products*. In physical space, *lengths* are the square roots of scalar products $\vec{\boldsymbol{n}}.\vec{\boldsymbol{n}}$, and *angles* are the inverse cosines of the scalar products of unit vectors in the $\vec{\boldsymbol{m}}$ and $\vec{\boldsymbol{n}}$ directions. In Hilbert space, *probabilities* are the squares of the moduli of scalar products $\langle v|u\rangle$ of normalized vectors, according to the Born Rule. Thus we expect the conservation of probability to derive from the conservation of angles in Hilbert space and from the *rigidity* of symmetry transformations in *physical* space being carried over to a similar rigidity in *Hilbert* space, because of the way Hilbert spaces are constructed using the Pythagorean construction. So a symmetry transformation in Hilbert space that takes a state $|n\rangle$ into a new state $|n'\rangle$ must preserve not only length, $\langle n|n\rangle = \langle n'|n'\rangle$, but also scalar products generally, $\langle m|n\rangle = \langle m'|n'\rangle$ (up to a phase factor). In order to preserve scalar products, the operator that effects the transformation must be not only *linear*, but also *unitary*. A *linear* operator $\hat{L}$ preserves the operation of scalar multiplication, $\hat{L}\lambda|\Psi\rangle = \lambda\hat{L}|\Psi\rangle$ (where $\lambda$ is a constant), and also preserves the operation of addition, $\hat{L}(|\Psi\rangle + |\Phi\rangle) = \hat{L}|\Psi\rangle + \hat{L}|\Phi\rangle$, the latter property implying that a *linear* operator is always the *same* operator, independent of the state vector it is operating on. These two aspects of linearity can be combined into one as

$$\hat{L}(\lambda_1|\Psi\rangle + \lambda_2|\Phi\rangle) = \lambda_1\hat{L}|\Psi\rangle + \lambda_2\hat{L}|\Phi\rangle \qquad \text{linearity} \qquad \text{(A23)}$$

Linearity enables a scalar product to be written $\langle m'|n'\rangle = \langle m|\hat{L}^\dagger\hat{L}|n\rangle$ (as opposed to $\langle m|\hat{\Omega}_m^\dagger\hat{\Omega}_n|n\rangle$ in the case of a non-linear operator $\hat{\Omega}_n$ that depends on the state $|n\rangle$ upon which it operates). But for scalar products to be conserved, the linear operator must also be *unitary*, $\hat{U}^\dagger = \hat{U}^{-1}$, so that $\langle m|\hat{U}^\dagger \hat{U}|n\rangle = \langle m|\hat{U}^{-1}\hat{U}|n\rangle = \langle m|n\rangle$. A transformation effected by such an operator is known as a *unitary transformation.* Indeed, Wigner proved in a famous theorem [14-18] that a symmetry transformation in Hilbert space that takes a state $|n\rangle$ into a new state $|n'\rangle$ is effected by a linear and unitary operator $\hat{U}(O', O)$ (or, in the case of time reversal, an antilinear and antiunitary operator), i.e.

$$|n'\rangle = \hat{U}(O', O)|n\rangle \quad \text{(A24a)}$$

$$|m'\rangle = \hat{U}(O', O)|m\rangle \quad \text{(A24b)}$$

where $\hat{U}(O', O)$ is independent of the state it operates on, but depends on the coordinate systems $O$ and $O'$ between which it effects the transformations (A24). It then follows from its unitarity that lengths, $\langle n|n\rangle = \langle n'|n'\rangle$, and scalar products generally, $\langle m|n\rangle = \langle m'|n'\rangle$, are preserved. Wigner's proof is lengthy, is based on the conservation of probabilities as the cardinal property of symmetry transformations and on the Born expression $\mathrm{P}_u(v) = |\langle v|u\rangle|^2$ for probabilities, and implicitly assumes a single Hilbert space.

In addition, the *covariance* of symmetry transformations is encapsulated in the *Haag-Wigner condition* [17,18] which states that the operators that effect the symmetry transformation of a QM state in Hilbert space must depend only on the *relation* between the two frames of reference in physical space-time, not on the intrinsic properties or absolute position in space-time of either frame, otherwise the homogeneity and isotropy of space-time would be violated. In other words if we represent the symmetry transformation of the state $|\Psi_O\rangle$ in the frame of observer $O$ into the state $|\Psi_{O'}\rangle$ in the frame of observer $O'$ as $|\Psi_{O'}\rangle = U(O', O)|\Psi_O\rangle$, the QM symmetry operator $U(O', O)$ that effects the transformation in Hilbert space is determined by the classical operator $T$ that takes frame $O$ into frame $O'$ in physical space-time, i.e. $|\Psi_{O'}\rangle = U(T)|\Psi_O\rangle$ [19]. This is consistent with the notion that the rigidity of a symmetry transformation $T$ in physical space-time is carried over to rigidity of the corresponding transformation in Hilbert space. The Haag-Wigner condition is basically a *relativity* condition: the parameters of the operator $T$ in physical space are *relative*, e.g. rotation operators in physical space depend on a *relative* angle $\alpha$, independent of absolute orientation, and this carries over to rotation through an angle $\omega(\alpha)$ in Hilbert space that is also purely relative, independent of absolute orientation.

The operators used to effect symmetry transformations are often written in *matrix* form, so it is important to clarify the difference between *unitary matrices* and *unitary operators*. Matrices $\mathbf{U}$ with the unitarity property $\mathbf{U}^{-1} = \mathbf{U}^\dagger$ (inverse equals Hermitian conjugate) are the complex analogues of orthogonal matrices $\mathbf{O}$, for which $\mathbf{O}^{-1} = \tilde{\mathbf{O}}$ (inverse equals transpose), and both orthogonal and unitary matrices have orthonormal rows and orthonormal columns. But not all *unitary* *matrices* are also *unitary operators*, because not all unitary matrices are also *linear* operators. The unitary property in a *matrix* should *not* be confused with the unitary property in a *transformation*: to reiterate, a *unitary transformation* is effected by a *unitary operator*, which is a *linear operator that is also unitary* [20]. Although linear operators in general are not necessarily also unitary, linear operators that operate on vectors representing physical states in a Hilbert space *are* automatically also unitary if we require that all vectors representing physical states in the Hilbert space are normalized: if $|n'\rangle = \hat{L}|n\rangle$ and we require $|n\rangle$ to be normalized, $\langle n|n\rangle = 1$, then $|n'\rangle$ will also be normalized, $\langle n'|n'\rangle = \langle n|\hat{L}^{\dagger}\hat{L}|n\rangle = \langle n|n\rangle = 1$, provided that $\hat{L}^{\dagger} = \hat{L}^{-1}$, i.e. that $\hat{L}$ is unitary.

So in summary, the Haag-Wigner condition requires a symmetry transformation operator to depend only on the *relation* between the two frames of reference $O$ and $O'$ it connects, independent of their absolute orientation. The simplest way to satisfy this condition is to require the symmetry operator to be *linear* (i.e. independent of what it operates on); and a linear operator in a Hilbert space in which all vectors representing physical states are required to be normalized will automatically also be *unitary*, thus ensuring that the transformation effected by the operator is *rigid*, as required of a symmetry transformation.

## Appendix 3 Applying the Haag-Wigner condition to the Arrow system

In Section 4 we noted that to fully incorporate the isotropy of space into the Arrow model we have to ensure that rotating the *entire experiment* through any angle $\alpha$ about the z-axis is a *symmetry transformation*. The *entire experiment* means the system (the Arrow itself along the $\vec{\boldsymbol{n}}$ axis) *plus* the measuring apparatus (the magnets along the $\vec{\boldsymbol{a}}$ axis at an angle $\theta$ to the Arrow, see for example Figure 1). Rotation of the entire experiment in physical space would mean rotating the Arrow vector $\vec{\boldsymbol{n}}$ clockwise by an angle $\alpha$ into a new vector $\vec{\boldsymbol{n}}'$, and at the same time rotating $\vec{\boldsymbol{a}}$ into $\vec{\boldsymbol{a}}'$ by the *same* angle $\alpha$. It is important not to confuse vectors $\vec{\boldsymbol{n}}_+$ in real physical space with the corresponding vectors $|n_+^A\rangle$ in Hilbert space: rotations of vectors in real space through a physical angle $\alpha$ correspond to rotations in Hilbert space through a *different* angle, the Pythagorean angle $\omega(\alpha)$. Note also that clockwise rotation of the Arrow $\vec{\boldsymbol{n}}$ in physical space corresponds to anticlockwise rotation of $|n_+^A\rangle$ in Hilbert space, simply because the $|a_+^A\rangle$ axis is taken to be the *x*-axis (see for example Figure 5). So the clockwise rotation $\vec{\boldsymbol{a}} \rightarrow \vec{\boldsymbol{a}}'$ by an angle $\alpha$ in physical space corresponds to an anticlockwise rotation $|n_+^A\rangle \rightarrow |n_+'^A\rangle$ by an angle $\omega(\alpha)$. It is also important to be clear which Hilbert space we are using: when the entire experiment is rotated, the variable A actually becomes a different variable A' that is incompatible with A (the Arrow can't be pointing along both $\vec{\boldsymbol{a}}$ and $\vec{\boldsymbol{a}}'$ simultaneously), so the two variables must be represented in two different Hilbert spaces $\mathcal{H}_A$ and $\mathcal{H}_{A'}$. In physical space, the rotation $\vec{\boldsymbol{n}} \to \vec{\boldsymbol{n}}'$ through an angle $\alpha$ is effected by a familiar type of orthogonal rotation matrix $\mathbf{R}(\alpha)$, but in the Hilbert space $\mathcal{H}_A$ the rotations $|n_+^A\rangle \to |n_+^A\rangle$ and $|n_-^A\rangle \to |n_-'^A\rangle$ through an angle $\omega(\alpha)$ would be effected by a matrix operator $\mathbf{Y^A}$.

We can ensure that $\mathbf{Y^A}$ is a *symmetry* operator by applying the Haag-Wigner condition (see Appendix 2). We begin by using equations (A6) to express all the relevant states in $\mathcal{H}_A$. The initial orientation $\vec{\boldsymbol{n}}$ of the Arrow is at an angle $\theta$ to $\vec{\boldsymbol{a}}$ (see Figure 1) giving

$$|n_+^A\rangle = f^{1/2}(\theta)|a_+^A\rangle + \left(1 - f(\theta)\right)^{1/2}|a_-^A\rangle \tag{A25a}$$

$$|n_-^A\rangle = -\left(1 - f(\theta)\right)^{1/2}|a_+^A\rangle + f^{1/2}(\theta)|a_-^A\rangle \tag{A25b}$$

The new orientation $\vec{\boldsymbol{n}}'$ of the system (the Arrow) is at an angle $\theta + \alpha$ to $\vec{\boldsymbol{a}}$, giving

$$|n_+'^A\rangle = f^{1/2}(\theta + \alpha)|a_+^A\rangle + \left(1 - f(\theta + \alpha)\right)^{1/2}|a_-^A\rangle \tag{A26a}$$

$$|n_-'^A\rangle = -\left(1 - f(\theta + \alpha\,)\right)^{1/2}|a_+^A\rangle + f^{1/2}(\theta + \alpha)|a_-^A\rangle \tag{A26b}$$

and the new orientation $\vec{\boldsymbol{a}}'$ of the measuring apparatus (the magnets) is at an angle $\alpha$ to $\vec{\boldsymbol{a}}$, giving corresponding new vectors in $\mathcal{H}_A$:

$$|a_+'^A\rangle = f^{1/2}(\alpha)|a_+^A\rangle + \left(1 - f(\alpha)\right)^{1/2}|a_-^A\rangle \tag{A27a}$$

$$|a_-'^A\rangle = -\left(1 - f(\alpha\,)\right)^{1/2}|a_+^A\rangle + f^{1/2}(\alpha)|a_-^A\rangle \tag{A27b}$$

In matrix notation, with the basis vectors of $\mathcal{H}_A$ written as

$$\boldsymbol{a}_+^A = \begin{pmatrix} 1 \\ 0 \end{pmatrix}, \ \boldsymbol{a}_-^A = \begin{pmatrix} 0 \\ 1 \end{pmatrix} \tag{A28}$$

we can write equations (A25), (A26) and (A27) as

$$\boldsymbol{n}_+^A = \begin{pmatrix} f^{1/2}(\theta) \\ \left(1 - f(\theta)\right)^{1/2} \end{pmatrix}, \qquad \boldsymbol{n}_-^A = \begin{pmatrix} -\left(1 - f(\theta)\right)^{1/2} \\ f^{1/2}(\theta) \end{pmatrix} \tag{A29a}$$

$$\boldsymbol{n}_+'^A = \begin{pmatrix} f^{1/2}(\theta + \alpha) \\ \left(1 - f(\theta + \alpha)\right)^{1/2} \end{pmatrix}, \quad \boldsymbol{n}_-'^A = \begin{pmatrix} -\left(1 - f(\theta + \alpha)\right)^{1/2} \\ f^{1/2}(\theta + \alpha) \end{pmatrix} \tag{A29b}$$

$$\boldsymbol{a}_+'^A = \begin{pmatrix} f^{1/2}(\alpha) \\ \left(1 - f(\alpha)\right)^{1/2} \end{pmatrix}, \qquad \boldsymbol{a}_-'^A = \begin{pmatrix} -\left(1 - f(\alpha)\right)^{1/2} \\ f^{1/2}(\alpha) \end{pmatrix} \tag{A29c}$$

which can be re-written using a matrix of the form

$$\mathbf{U}(x) = \begin{pmatrix} f^{1/2}(x) & -\left(1 - f(x)\right)^{1/2} \\ \left(1 - f(x)\right)^{1/2} & f^{1/2}(x) \end{pmatrix} \tag{A30}$$

to give

$$\boldsymbol{n}_+^A = \mathbf{U}(\theta)\boldsymbol{a}_+^A, \qquad \boldsymbol{n}_-^A = \mathbf{U}(\theta)\boldsymbol{a}_-^A \tag{A31a}$$

$$\boldsymbol{n}_+'^A = \mathbf{U}(\theta + \alpha)\boldsymbol{a}_+^A, \qquad \boldsymbol{n}_-'^A = \mathbf{U}(\theta + \alpha)\boldsymbol{a}_-^A \tag{A31b}$$

$$\boldsymbol{a}_+'^A = \mathbf{U}(\alpha)\boldsymbol{a}_+^A, \qquad \boldsymbol{a}_-'^A = \mathbf{U}(\alpha)\boldsymbol{a}_-^A \tag{A31c}$$

and we see that $\mathbf{U}(x)$ is a *unitary* matrix because it has orthonormal rows and columns. Actually, we only need to consider the transformations $\boldsymbol{a}_+^A \rightarrow \boldsymbol{a}_+'^A$ and $\boldsymbol{n}_+^A \rightarrow \boldsymbol{n}_+'^A$ in what follows: the transformations $\boldsymbol{a}_-^A \rightarrow \boldsymbol{a}_-'^A$ and $\boldsymbol{n}_-^A \rightarrow \boldsymbol{n}_-'^A$ follow similarly.

We seek to find a matrix operator $\mathbf{Y^A}(\alpha;\phi)$ that effects the transformation $\boldsymbol{n}_+^A \rightarrow \boldsymbol{n}_+'^A$ in Hilbert space $\mathcal{H}_A$ corresponding to the rotation $\vec{\boldsymbol{n}} \rightarrow \vec{\boldsymbol{n}}'$ in real physical space through an angle $\alpha$ from an initial *absolute* (i.e. lab frame) orientation $\phi$ of $\vec{\boldsymbol{n}}$ to a final absolute orientation $\phi + \alpha$ of $\vec{\boldsymbol{n}}'$. We take the absolute (lab frame) orientation in physical space of $\vec{\boldsymbol{a}}$ to be $\phi = 0$, so the absolute orientation of $\vec{\boldsymbol{n}}$ becomes $\phi = \theta$, and in Hilbert space $\mathcal{H}_A$ we now have

$$\boldsymbol{a}_+'^A = \mathbf{Y^A}(\alpha;0)\boldsymbol{a}_+^A \tag{A32}$$

and

$$\boldsymbol{n}_+'^A = \mathbf{Y^A}(\alpha;\theta)\boldsymbol{n}_+^A \tag{A33}$$

The matrices $\mathbf{Y^A}(\alpha;\theta)$ and $\mathbf{Y^A}(\alpha;0)$ are not in general the same, because in the absence of symmetry the transformation matrix $\mathbf{Y^A}(\alpha;\phi)$ may depend on the absolute orientation $\phi$ of the vector upon which $\mathbf{Y^A}$ operates. But the Haag-Wigner condition tells us that if $\mathbf{Y^A}$ is a matrix that induces a *symmetry* transformation, it can only depend upon the *relation* of the initial and transformed vectors, and not on the absolute orientation of either one. This means that $\mathbf{Y^A}(\alpha;\phi)$ must be independent of the starting angle $\phi$, so the Haag-Wigner condition can be stated as

$$\mathbf{Y^A}(\alpha;\theta) = \mathbf{Y^A}(\alpha;0) \equiv \mathbf{Y^A}(\alpha) \tag{A34}$$

This basically says that the *same* matrix $\mathbf{Y^A}(\alpha)$ is used to rotate both $\boldsymbol{n}_+^A$ and $\boldsymbol{a}_+^A$ by the *same* angle, thus giving a *rigid* rotation of the system in the Hilbert space $\mathcal{H}_A$, as required for a symmetry operator: as noted above, the simplest way to satisfy the Haag-Wigner condition is to require the matrix $\mathbf{Y^A}(\alpha)$ to be a *linear* operator (i.e. independent of what it operates on), and since it operates on normalized vectors in Hilbert spaces it is also *unitary* (scalar product preserving).

But what is the form of the operator $\mathbf{Y^A}(\alpha)$? We see by comparing (A32) and (A31c) that

$$\mathbf{Y^A}(\alpha;0) = \mathbf{U}(\alpha) \tag{A35}$$

From (A31) we have $\boldsymbol{n}_+'^A = \mathbf{U}(\theta+\alpha)\boldsymbol{a}_+^A$ and $\boldsymbol{n}_+^A = \mathbf{U}(\theta)\boldsymbol{a}_+^A$, and we use the unitarity of the matrix $\mathbf{U}(x)$ to write

$$\boldsymbol{a}_+^A = \mathbf{U}^{-1}(\theta)\boldsymbol{n}_+^A = \mathbf{U}^{\dagger}(\theta)\boldsymbol{n}_+^A \tag{A36}$$

giving

$$\boldsymbol{n}_+'^A = \mathbf{U}(\theta+\alpha)\mathbf{U}^{\dagger}(\theta)\boldsymbol{n}_+^A \tag{A37}$$

so that

$$\mathbf{Y^A}(\alpha;\theta) = \mathbf{U}(\theta+\alpha)\mathbf{U}^{\dagger}(\theta) \tag{A38}$$

The matrix $\mathbf{Y}^{\mathbf{A}}(\alpha;\theta)$ in equation (A38) must be a unitary matrix because it is the product of two unitary matrices, and it is easily shown that the product of two unitary matrices is also unitary. But although $\mathbf{Y}^{\mathbf{A}}(\alpha;\theta)$ is a unitary *matrix*, it is not in general a *linear and unitary operator*, because it depends on $\theta$. However, if we now require that the Haag-Wigner condition (A34) be satisfied in the Hilbert space $\mathcal{H}_A$, we obtain

$$\mathbf{Y}^{\mathbf{A}}(\alpha;\theta) = \mathbf{U}(\theta+\alpha)\mathbf{U}^{\dagger}(\theta) = \mathbf{U}(\alpha) \tag{A39}$$

which shows that $\mathbf{Y}^{\mathbf{A}}(\alpha;\theta)$ *is* now linear, unitary, and independent of $\theta$. But this condition now imposes rather stringent restrictions on the form of the probability function $f$, because $\mathbf{U}(\theta+\alpha)\mathbf{U}^{\dagger}(\theta)$ depends in general on $\theta$, whereas $\mathbf{U}(\alpha)$ does not.

We now use the Pythagorean construction from equation (A1) to write the probability function $f(\alpha)$ as

$$f(\alpha) = \cos^2\omega(\alpha) \tag{A40}$$

where $\alpha$ is the actual physical angle in space through which the Arrow is rotated, while $\omega(\alpha)$ is the Pythagorean angle in Hilbert space $\mathcal{H}_A$. The connection between $\alpha$ and $\omega(\alpha)$ may be very complicated in the general case, but the linearity condition (A40) imposed by symmetry requires, as we now show, a very simple and unique relationship between $\alpha$ and $\omega(\alpha)$. Using (A40) in (A30) now gives

$$\mathbf{U}(\alpha) = \begin{pmatrix} \cos\omega(\alpha) & -\sin\omega(\alpha) \\ \sin\omega(\alpha) & \cos\omega(\alpha) \end{pmatrix} \tag{A41}$$

for the right-hand side of (A39), and the left-hand side of (A40) now becomes

$$\begin{aligned} \mathbf{U}(\theta+\alpha)\mathbf{U}^{\dagger}(\theta) &= \begin{pmatrix} \cos\omega(\theta+\alpha) & -\sin\omega(\theta+\alpha) \\ \sin\omega(\theta+\alpha) & \cos\omega(\theta+\alpha) \end{pmatrix} \begin{pmatrix} \cos\omega(\theta) & \sin\omega(\theta) \\ -\sin\omega(\theta) & \cos\omega(\theta) \end{pmatrix} \\ &= \begin{pmatrix} \cos[\omega(\theta+\alpha)-\omega(\theta)] & -\sin[\omega(\theta+\alpha)-\omega(\theta)] \\ \sin[\omega(\theta+\alpha)-\omega(\theta)] & \cos[\omega(\theta+\alpha)-\omega(\theta)] \end{pmatrix} \end{aligned} \tag{A42}$$

Using (A39) to equate (A42) and (A41) now gives the Haag-Wigner condition as

$$\cos[\omega(\theta+\alpha)-\omega(\theta)] = \cos\omega(\alpha) \tag{A43a}$$

and

$$\sin[\omega(\theta+\alpha)-\omega(\theta)] = \sin\omega(\alpha) \tag{A43b}$$

or

$$\omega(\theta+\alpha)-\omega(\theta) = \omega(\alpha) \tag{A44}$$

which says that the angle of rotation in Hilbert space, $\omega(\alpha)$, is independent of the absolute orientation $\theta$ of the Arrow in physical space. This condition can be satisfied only if $\omega(\alpha)$ is linearly related to $\alpha$, i.e. if $\omega(\alpha)$ is just $\alpha$ multiplied by a constant $c$:

$$\omega(\alpha) = c\alpha \tag{A45}$$

To determine the constant $c$ we note that in deriving the matrix $\mathbf{Y}^{\mathbf{A}}(\alpha)$ the only restriction we have imposed on the probability function $f(\theta)$ for the Arrow is the normalization of probabilities, $f(0) + f(\pi) = 1$, but we have not yet applied the exclusivity conditions $f(0) = 1$ and $f(\pi) = 0$ (which simply state that if the Arrow is definitely pointing along $\vec{\boldsymbol{a}}_+$ it cannot also be pointing along $\vec{\boldsymbol{a}}_-$). Applying these conditions gives $f(0) = \cos^2 \omega(0) = 1$, implying $\omega(0) = 0$, and $f(\pi) = \cos^2 \omega(\pi) = 0$ , implying $\omega(\pi) = \pi/2$ , which shows that the constant in (A45) must be ½ , so we have the very simple relationship

$$\omega = ½\alpha \tag{A46}$$

between the physical angle $\alpha$ in the Arrow model and the corresponding Pythagorean angle $\omega$ in Hilbert space when symmetry restrictions from the isotropy of space are imposed. This gives our final results

$$f(\alpha) = \cos^2 ½\alpha \tag{A47}$$

and

$$\mathbf{U}(\alpha) = \begin{pmatrix} \cos ½\alpha & -\sin ½\alpha \\ \sin ½\alpha & \cos ½\alpha \end{pmatrix} = \mathbf{Y}^{\mathbf{A}}(\alpha) \tag{A48}$$

for our rotation operator. This is precisely the result for the quantum mechanics of a spin ½ system – and yet we have just shown that this result emerges for a *classical* probabilistic system possessing incompatible variables (our Arrow) when we apply symmetry constraints!

Our linear and unitary matrix operator $\mathbf{Y}^{\mathbf{A}}(\alpha)$ in (A48) operates in the Hilbert space $\mathcal{H}_A$. Would we get an operator of the same form if we worked in a different space, for example $\mathcal{H}_{A'}$? The answer is yes! To obtain the operator $\mathbf{Y}^{\mathbf{A}'}(\alpha)$ we follow the same procedure that we used to derive $\mathbf{Y}^{\mathbf{A}}(\alpha)$, but in $\mathcal{H}_{A'}$ instead of $\mathcal{H}_A$. Working in $\mathcal{H}_{A'}$ gives us an opportunity to pause to examine more closely the conservation of probabilities under symmetry transformations. Recall that when the entire experiment is rotated the variable A actually becomes a different variable A', so before the rotation we are considering $\mathrm{P}_{(n_+)}(a_+)$, and after the rotation we are considering $\mathrm{P}_{(n'_+)}(a'_+)$. Since A and A' are incompatible variables, $\mathrm{P}_{(n_+)}(a_+)$ must be evaluated in Hilbert space $\mathcal{H}_A$ while $\mathrm{P}_{(n'_+)}(a'_+)$ must be evaluated in Hilbert space $\mathcal{H}_{A'}$ . For the results of experiments to be the same before and after the rotation we therefore require

$$\mathrm{P}_{(n_+)}(a_+) = \mathrm{P}_{(n'_+)}(a'_+) \tag{A49}$$

where

$$\mathrm{P}_{(n_+)}(a_+) = |\langle a^A_+ | n^A_+ \rangle|^2 \tag{A50a}$$

$$\mathrm{P}_{(n'_+)}(a'_+) = \left|\langle a'^{A'}_+ | n'^{A'}_+ \rangle\right|^2 \tag{A50b}$$

giving the requirement for probability conservation as

$$\langle a^A_+ | n^A_+ \rangle = \langle a'^{A'}_+ | n'^{A'}_+ \rangle \tag{A51}$$

Symmetry transformations are *rigid* (angle-preserving) in real physical space, e.g. the rotated Arrow vector $\vec{\boldsymbol{n}}'$ is at the same angle $\theta$ to $\vec{\boldsymbol{a}}'$ as $\vec{\boldsymbol{n}}$ is to $\vec{\boldsymbol{a}}$, and to satisfy the conservation of probabilities in equation (A51) the corresponding angles in Hilbert space must also stay the same during the transformation, meaning that $|n'^{A'}_{+}\rangle$ must be at the same angle $\omega(\theta)$ to $|a'^{A'}_{+}\rangle$ as $|n^{A}_{+}\rangle$ is to $|a^{A}_{+}\rangle$. To achieve this, we simply *construct* the vectors in $\mathcal{H}_A$ and $\mathcal{H}_{A'}$ using the probability functions $f(\theta)$ of the angle $\theta$ in physical space, as in the Pythagorean construction of equation (A6), giving

$$|n^{A}_{+}\rangle = f^{1/2}(\theta)|a^{A}_{+}\rangle + \left(1 - f(\theta)\right)^{1/2}|a^{A}_{-}\rangle \quad \text{(A52a)}$$

$$|n'^{A'}_{+}\rangle = f^{1/2}(\theta)|a'^{A'}_{+}\rangle + \left(1 - f(\theta)\right)^{1/2}|a'^{A'}_{-}\rangle \quad \text{(A52b)}$$

from which it is clear that (A51) is satisfied. The *rigidity* of the symmetry transformation $\vec{\boldsymbol{n}} \to \vec{\boldsymbol{n}}'$ thus carries over from real physical space to Hilbert space. So the conservation of probabilities in a symmetry transformation follows automatically from the (Pythagorean) way the Hilbert spaces $\mathcal{H}_A$ and $\mathcal{H}_{A'}$ are constructed from the probability function $f(\theta)$.

However, it is clear that the conservation of probabilities tells us nothing about the *form* of the probability function $f(\theta)$, because (A51) is satisfied regardless of the form of $f(\theta)$. To get the *form* (A48) of $f(\theta)$ we had to apply the Haag-Wigner condition (A39), which required that the matrix $\mathbf{Y}^{\mathbf{A}}(\alpha;\phi)$ for rotation through an angle $\omega(\alpha)$ in the Hilbert space $\mathcal{H}_A$ be independent of the starting angle $\phi$. We now derive the transformation matrix $\mathbf{Y}^{\mathbf{A}'}(\alpha)$ in $\mathcal{H}_{A'}$, starting with expressions analogous to equations (A25) and (A26), but for the relevant states in $\mathcal{H}_{A'}$ instead of $\mathcal{H}_A$ (noting that in physical space the Arrow $\vec{\boldsymbol{n}}$ is at an angle $\theta - \alpha$ to $\vec{\boldsymbol{a}}'$, and $\vec{\boldsymbol{n}}'$ is at an angle $\theta$ to $\vec{\boldsymbol{a}}'$)

$$|n^{A'}_{+}\rangle = f^{1/2}(\theta - \alpha)|a'^{A'}_{+}\rangle + \left(1 - f(\theta - \alpha)\right)^{1/2}|a'^{A'}_{-}\rangle \quad \text{(A53a)}$$

$$|n'^{A'}_{+}\rangle = f^{1/2}(\theta)|a'^{A'}_{+}\rangle + \left(1 - f(\theta)\right)^{1/2}|a'^{A'}_{-}\rangle \quad \text{(A53b)}$$

and thence obtain, analogous to equations (A31),

$$\boldsymbol{n}^{A'}_{+} = \mathbf{U}(\theta - \alpha)\boldsymbol{a}'^{A'}_{+} \quad \text{(A54a)}$$

$$\boldsymbol{n}'^{A'}_{+} = \mathbf{U}(\theta)\boldsymbol{a}'^{A'}_{+} \quad \text{(A54b)}$$

$$\boldsymbol{a}'^{A'}_{+} = \mathbf{U}(\alpha)\, \boldsymbol{a}^{A'}_{+} \quad \text{(A54c)}$$

which gives, analogous to (A35) and (A38),

$$\mathbf{Y}^{\mathbf{A}'}(\alpha;0) = \mathbf{U}(\alpha) \quad \text{(A55a)}$$

$$\mathbf{Y}^{\mathbf{A}'}(\alpha;\theta) = \mathbf{U}(\theta)\mathbf{U}^{\dagger}(\theta - \alpha) \quad \text{(A55b)}$$

still taking the absolute orientation in physical space of $\vec{\boldsymbol{a}}$ to be $\phi = 0$. Imposing the Haag-Wigner requirement that $\mathbf{Y}^{\mathbf{A}'}(\alpha;\phi)$ is independent of $\phi$ then leads to the condition

$$\cos[\omega(\theta) - \omega(\theta - \alpha)] = \cos \omega(\alpha) \qquad \text{(A56)}$$

analogous to (A41), which is only possible if $\omega(\alpha) = c\alpha$, giving again $\omega = ½\alpha$ as in (A46), and thence

$$\mathbf{Y}^{\mathbf{A}'}(\alpha) = \begin{pmatrix} \cos ½\alpha & -\sin ½\alpha \\ \sin ½\alpha & \cos ½\alpha \end{pmatrix} = \mathbf{Y}^{\mathbf{A}}(\alpha) \qquad \text{(A57)}$$

so the operator $\mathbf{Y}(\alpha)$ is indeed the same in both $\mathcal{H}_A$ and $\mathcal{H}_{A'}$. But $\mathbf{Y}(\alpha)$ is the same in both Hilbert spaces *only* after applying the Haag-Wigner condition for it to be a symmetry operator for all $\alpha$, i.e. *only* if the Arrow experiment is carried out in a space-time that is isotropic about the z-axis.

In summary we have shown that introducing symmetry constraints into the Arrow model led to a unique form for the probability function, $f(\theta) = \cos^2 ½\theta$, equation (A47). The symmetry constraint we used was full rotational symmetry, i.e. isotropy of space, which required $f_A$, $f_B$, $f_C$… to be all the *same* function $f$ (equation (40)) and also required the matrix operator $\mathbf{Y}(\alpha)$ that effects rotations in Hilbert space to satisfy the Haag-Wigner condition, which led to a unique form that is both linear and unitary, equation (A57). We obtained the same unique results for $f$ and $\mathbf{Y}(\alpha)$ using either Hilbert space $\mathcal{H}_A$ or Hilbert space $\mathcal{H}_{A'}$ for the derivation. Clearly the same argument can be applied to any of the other Hilbert spaces $\mathcal{H}_B$, $\mathcal{H}_C$..... for the variables B, C…., and the results for $f$ and $\mathbf{Y}(\alpha)$ will be the same in each space. All of the Hilbert spaces $\mathcal{H}_A, \mathcal{H}_B, \mathcal{H}_C$..... therefore become equivalent, so one should be able to calculate the probabilities of values of A, B, C.... in any one of the spaces, i.e.

$$\mathrm{P}_{(n_i)}(a_j) = \left|\langle a_j^A | n_i^A \rangle\right|^2 = \left|\langle a_j^B | n_i^B \rangle\right|^2 = \left|\langle a_j^C | n_i^C \rangle\right|^2 = \cdots \qquad \text{(A58)}$$

Consider, for example, the probability $\mathrm{P}_{(n_+)}(a_+)$ which in general would be calculated in $\mathcal{H}_A$ because it involves a measurement of the variable A, and is given by

$$\mathrm{P}_{(n_+)}(a_+) = |\langle a_+^A | n_+^A \rangle|^2 = f(\theta_n) = \cos^2 ½\theta_n \qquad \text{(A59)}$$

where $\theta_n$ is the angle between the vectors $\vec{\boldsymbol{a}}$ and $\vec{\boldsymbol{n}}$. and we have used equation (A48) to express our final result (A61) for Hilbert space $\mathcal{H}_A$. For comparison, we now calculate the quantity $|\langle a_+^B | n_+^B \rangle|^2$, which in the general case (no symmetry) would *not* be equal to $|\langle a_+^A | n_+^A \rangle|^2$ and hence could *not* be used to calculate the probability $\mathrm{P}_{(n_+)}(a_+)$. If the angle between $\vec{\boldsymbol{a}}$ and $\vec{\boldsymbol{b}}$ in physical space is $\theta_b$, the angle between $\vec{\boldsymbol{n}}$ and $\vec{\boldsymbol{b}}$ then becomes $\theta_b - \theta_n$, so we then have

$$|a_+^B\rangle = f^{1/2}(\theta_b)|b_+^B\rangle + \left(1 - f(\theta_b)\right)^{1/2}|b_-^B\rangle \qquad \text{(A60a)}$$

$$|n_+^{\mathrm{B}}\rangle = f^{1/2}((\theta_b - \theta_n)|b_+^B\rangle + (1 - f(\theta_b - \theta_n))^{1/2}|b_-^B\rangle \qquad \text{(A60b)}$$

Hence, using the orthonormality of the basis set $\{|b_+^B\rangle, |b_-^B\rangle\}$, we find

$$\langle a_+^B | n_+^B \rangle = f^{1/2}(\theta_b) f^{1/2}(\theta_b - \theta_n) + (1 - f(\theta_b))^{1/2}(1 - f(\theta_b - \theta_n))^{1/2} \qquad \text{(A62)}$$

which shows that in the general case, when $f(\theta) \neq \cos^2 ½\theta$, we would find that $\langle a_+^B | n_+^B \rangle \neq \langle a_+^A | n_+^A \rangle$. But when symmetry is present we have $f(\theta) = \cos^2 ½\theta$ and thence

$$\langle a_+^B | n_+^B \rangle = \cos ½\theta_b \cos ½(\theta_b - \theta_n) + \sin ½\theta_b \sin ½(\theta_b - \theta_n) = \cos ½\theta_n \quad \text{(A63)}$$

giving

$$|\langle a_+^B | n_+^B \rangle|^2 = |\langle a_+^A | n_+^A \rangle|^2 = \cos^2 ½\theta_n = \mathrm{P}_{(n_+)}(a_+) \quad \text{(A64)}$$

It is thus clear that when symmetry is present we can select any of the Hilbert spaces $\mathcal{H}_A, \mathcal{H}_B, \mathcal{H}_C$..... to calculate the probability $\mathrm{P}_{(n_+)}(a_+)$. Furthermore it is easily shown that any of these Hilbert spaces can be used to calculate probabilities of outcomes for the measurement of *any* of the variables A, B, C…. Therefore in the Arrow system, as in QM, only a single Hilbert space is necessary, but *only* when *symmetry* is present, specifically not only the $C_2$ symmetry of the Arrow itself, but also the isotropy of space.

## Appendix 4 The equivalence assumption

Two representations $\Gamma_\varphi$ and $\Gamma_{\varphi'}$ based respectively on two different sets of basis vectors $\{\boldsymbol{\varphi_i}\}$ and $\{\boldsymbol{\varphi'_i}\}$, are defined as *equivalent* if the corresponding matrix representatives $\boldsymbol{\Gamma_\varphi}(\mathrm{X})$ and $\boldsymbol{\Gamma_{\varphi'}}(\mathrm{X})$ for each symmetry operation X of the group are related by a similarity transformation

$$\boldsymbol{\Gamma_{\varphi'}}(\mathrm{X}) = \mathbf{S}\, \boldsymbol{\Gamma_\varphi}(\mathrm{X})\, \mathbf{S^{-1}} \quad \text{(A65)}$$

as in equation (74) of Section 5, where the matrix $\mathbf{S}$ is the matrix that relates individual basis vectors $\boldsymbol{\varphi_i}$ of $\Gamma_\varphi$ to the corresponding individual basis vectors $\boldsymbol{\varphi'_i}$ of $\Gamma_{\varphi'}$

$$\boldsymbol{\varphi'_i} = \mathbf{S}\boldsymbol{\varphi_i} \quad \text{(A66)}$$

as in equation (73). In standard QM the basis sets $\{\boldsymbol{\varphi_i}\}$ and $\{\boldsymbol{\varphi'_i}\}$ are in the *same* Hilbert space, but in our GIV theory the two basis sets are in *different* Hilbert spaces, so we need to justify our assumption that the representations $\Gamma_A$ in Hilbert space $\mathcal{H}_A$ and $\Gamma_B$ in Hilbert space $\mathcal{H}_B$, related by the *generalized* similarity transformation

$$\boldsymbol{\Gamma_B} = \mathbf{S_{AB}}\boldsymbol{\Gamma_A}\mathbf{S_{AB}^{-1}} \quad \text{(A67)}$$

as in equation (74), can still be considered *equivalent*. This *generalized* equivalence assumption can be justified as follows. In addition to the *fundamental* variables A, B, C,… a system may also possess *derived* variables that are functions of the fundamental variables. The most important of these turn out to be variables, known as Casimir invariants, that are compatible with *all* of the fundamental variables. A familiar example from the full rotation group $R_3$ is the derived variable $J^2 = J_x^2 + J_y^2 + J_z^2$, which is compatible with $J_x, J_y$ and $J_z$ and has the possible values $J(J+1)$ where $J$ is an integer or half-integer. Casimir operators are represented by constant multiples of the unit matrix **1**, so that they commute with all other elements of the group, e.g. $J(J+1)\mathbf{1}$ in the full rotation group, and spin s**1** and rest mass m**1** in the Poincaré group.

According to Schur's lemma [19], the Casimir invariants serve to distinguish between *different* (i.e. *inequivalent*) irreps, which are therefore identified and labelled with their Casimir invariants, e.g. J in the full rotation group, and spin and rest mass in the Poincaré group. Hence, for a given type of elementary system, characterized by its spin and rest mass, all of its irreps must be equivalent. The same must be true for an elementary GIV system: if the irreps $\Gamma_A, \Gamma_B, \Gamma_C$ ... in different Hilbert spaces were *not equivalent*, that would imply they had different Casimir invariants depending on the variable A, B, or C…. being measured. For example, if A, B and C were an elementary system's spin variables in the $x, y$ and $z$ directions, the system could behave like a spin 0 particle in the $x$ direction, like a spin ½ particle in the $y$ direction, and like a spin 1 particle in the $z$ direction. Not only has this type of behaviour never been observed experimentally, but, more fundamentally, such behaviour would violate the symmetry of spacetime. Similarly, the observed rest mass must always be the same, irrespective of which variable A, B or C is being measured. We must therefore require *equal Casimir invariants* in the different Hilbert spaces, which means the irreps $\Gamma_A, \Gamma_B, \Gamma_C \ldots$ in the different Hilbert spaces must therefore be *equivalent.*

## References

1. Kirkpatrick, K.A. Quantal behavior in classical probability. *Found. Phys. Lett*. **2003**, *16*, 199-224.
2. Hegstrom, R.A.; Adshead, G. Incompatible variables and "quantal" phenomena in psychology. *J. North Carolina Acad. Sci.* **2011**, *127*, 18-27.
3. Weinberg, S. *Elementary particles and the laws of physics. The 1986 Dirac Memorial Lectures*; Cambridge University Press, UK, 1987.
4. Sakurai, J. J. *Modern Quantum Mechanics*; Benjamin Cummings: San Francisco, CA, USA, 1985.
5. Hughes, R.I.G. *The structure and interpretation of quantum mechanics*; Harvard University Press: Cambridge, MA, USA, 1989.
6. Wilczek, F. *Fantastic Realities*; World Scientific Publishing: Singapore, 2006.
7. Bohr, A.; Ulfbeck, O. Primary manifestation of symmetry. Origin of quantal indeterminacy. *Rev. Mod. Phys.* **1995**, *67*, 1-35.
8. Gleason, A.M. Measures on the closed subspaces of a Hilbert space. *J. Math. and Mech.* **1957**, *6*, 885-893.
9. Weinberg, S. *Lectures on Quantum Mechanics*, 2nd ed.; Cambridge University Press: Cambridge, UK, 2015.
10. Deutsch, D. Quantum theory of probability and decisions. *Proc. R. Soc. London A* **1999,** *455*, 3129-3137.

11. Wallace, D. Everettian rationality: defending Deutsch's approach to probability in the Everett interpretation. **2003**, *34,* 415-438.
12. Zurek, W. H. Probabilities from entanglement, Born's rule $p_k = |\psi_k|^2$ from envariance. *Phys. Rev. A* **2005**, *71*, 052105.
13. Brumer, P.; Gong, J. Born rule in quantum and classical mechanics. *Phys. Rev. A* **2006**, *73*, 052109.
14. Weinberg, S. *The Quantum Theory of Fields I*;. Cambridge University Press, Cambridge, UK, 1995.
15. Cheng, T-P. *Relativity, Gravitation and Cosmology*; Oxford University Press, Oxford, UK, 2010.
16. Melvin, M.A. Elementary Particles and Symmetry Principles. *Rev. Mod. Phys.* **1960**, *32*, 477-485.
17. Wigner, E. P. *Group Theory and its applications to the Quantum Mechanics of Atomic Spectra*; Academic Press: New York, NY, USA (1959).
18. Wigner, E.P. Relativistic invariance in quantum mechanics. *Nuovo Cimento* **1956**, *3,* 517-532.
19. Schweber, S. S. *An Introduction to Relativistic Quantum Field Theory*; Harper & Row, New York, NY, USA, 1962.
20. Jordan, T. F. *Linear Operators for Quantum Mechanics*; Dover: Mineola, NY, USA, 1997.
21. Atkins, P.W. and Friedmann, R.S. *Molecular Quantum Mechanics*, 5th ed.; Oxford University Press, Oxford, UK, 2011.
22. Arfken, G. B., Weber, H. J. and Harris, *F. E. Mathematical Methods for Physicists*, 7th ed.; Academic Press, Elsevier, New York, NY, USA, 2012.
23. Hecht, E. *Optics*, 4th ed.; Pearson/Addison-Wesley: Boston, MA, USA, 2002.
24. Feynman, R. P. and Hibbs, A. R. *Quantum Mechanics and Path Integrals*, emended edition; Dover: Mineola, NY, USA, 2010.
25. Born, M. Zur Quantenmechanik der Stossvorgänge. *Z. Phys.* **1926**, *37*, 863-867.
26. Pleinert, M-O.; von Zanthier, J.; Lutz, E.: Many-particle interference to test Born's Rule. *Phys. Rev. Res*. **2020**, *2*, 012051 (R).
27. Koopman, B.O. Hamiltonian systems and transformation in Hilbert spaces. *Proc. Nat. Acad. Sci. USA* **1931**, *17*, 315-318.
28. Cooke, R.; Keane, M.; Moran, W. An elementary proof of Gleason's theorem. *Math. Proc. Cambridge Phil. Soc.* **1985**, *98*, 117-128.